\documentclass{emulateapj}

\sfcode`A=1000 \sfcode`B=1000 \sfcode`C=1000 \sfcode`D=1000
\sfcode`E=1000 \sfcode`F=1000 \sfcode`G=1000 \sfcode`H=1000
\sfcode`I=1000 \sfcode`J=1000 \sfcode`K=1000 \sfcode`L=1000
\sfcode`M=1000 \sfcode`N=1000 \sfcode`O=1000 \sfcode`P=1000
\sfcode`Q=1000 \sfcode`R=1000 \sfcode`S=1000 \sfcode`T=1000
\sfcode`U=1000 \sfcode`V=1000 \sfcode`W=1000 \sfcode`X=1000
\sfcode`Y=1000 \sfcode`Z=1000

\newcommand{\tf}{$\rm{TFIT}_{\rm{CANDELS+IB}}$}

\shorttitle{PHOTO-$z$ in ECDFS}
\shortauthors{Hsu et al.}

\usepackage{booktabs}
\usepackage{multirow}
\usepackage{graphicx}

\usepackage{natbib}
\begin{document}


\title{CANDELS/GOODS-S, CDFS, ECDFS: Photometric Redshifts For Normal
  and for X-Ray-Detected Galaxies} 

\author{Li-Ting Hsu\altaffilmark{1}, Mara Salvato\altaffilmark{1},
  Kirpal Nandra\altaffilmark{1}, Marcella Brusa\altaffilmark{1,2,15},
  Ralf Bender\altaffilmark{1}, Johannes Buchner\altaffilmark{1}, 
  Jennifer L. Donley\altaffilmark{3}, Dale D. Kocevski\altaffilmark{4}, 
  Yicheng Guo\altaffilmark{5,6}, Nimish P. Hathi\altaffilmark{7}, 
  Cyprian Rangel\altaffilmark{8}, 
  S. P. Willner\altaffilmark{9},
  Murray Brightman\altaffilmark{1}, 
  Antonis Georgakakis\altaffilmark{1}, 
  Tam\'{a}s Budav\'{a}ri\altaffilmark{12}, Alexander S. Szalay\altaffilmark{12}, 
  Matthew L. N. Ashby\altaffilmark{9}, Guillermo Barro\altaffilmark{5},
  Tomas Dahlen\altaffilmark{10},
  Sandra M. Faber\altaffilmark{5}, Henry C. Ferguson\altaffilmark{10}, 
  Audrey Galametz\altaffilmark{11},
  Andrea Grazian\altaffilmark{11}, Norman A. Grogin\altaffilmark{10},
  Kuang-Han Huang\altaffilmark{16}, 
  Anton M. Koekemoer\altaffilmark{10}, Ray A. Lucas\altaffilmark{10},
  Elizabeth McGrath\altaffilmark{13}, Bahram Mobasher\altaffilmark{14}, 
  Michael Peth\altaffilmark{12}, David J. Rosario\altaffilmark{1}, Jonathan R. Trump\altaffilmark{5}}

\altaffiltext{1}{Max-Planck-Institut f\"ur extraterrestrische Physik, Giessenbachstrasse D-85748 Garching, Germany}
\altaffiltext{2}{Dipartimento di Fisica e Astronomia, Universit\`{a} di Bologna, viale Berti Pichat 6/2, 40127 Bologna, Italy}
\altaffiltext{3}{Los Alamos National Laboratory, Los Alamos, NM, USA}
\altaffiltext{4}{Department of Physics and Astronomy, University of Kentucky, Lexington, KY 40506, USA}
\altaffiltext{5}{UCO/Lick Observatory, Department of Astronomy and Astrophysics, University of California, Santa Cruz, CA, USA}
\altaffiltext{6}{Department of Astronomy, University of Massachusetts, Amherst, MA, USA}
\altaffiltext{7}{Aix Marseille Universit\'{e}, CNRS, LAM (Laboratoire d'Astrophysique de Marseille) UMR 7326, 13388, Marseille, France}
\altaffiltext{8}{Astrophysics Group, Imperial College London, Blackett Laboratory, Prince Consort Road, London SW7 2AZ}
\altaffiltext{9}{Harvard-Smithsonian Center for Astrophysics, Cambridge, MA, USA}
\altaffiltext{10}{Space Telescope Science Institute, Baltimore, MD, USA}
\altaffiltext{11}{INAF-Osservatorio di Roma, I-00040 Monteporzio, Italy}
\altaffiltext{12}{Department of Physics and Astronomy, The Johns Hopkins University, Baltimore, MD, USA}
\altaffiltext{13}{Department of Physics and Astronomy, Colby College, Waterville, ME, USA}
\altaffiltext{14}{Department of Physics and Astronomy, University of California, Riverside, CA, USA}
\altaffiltext{15}{INAF-Osservatorio Astronomico di Bologna, via Ranzani 1, 40127 Bologna, Italy}
\altaffiltext{16}{Department of Physics, University of California Davis, CA, USA}

\begin{abstract}
   We present  photometric redshifts and
    associated probability distributions for all detected sources in
    the Extended Chandra Deep Field South (ECDFS). The work makes
    use of the most up-to-date data from the Cosmic 
    Assembly Near-IR Deep Legacy Survey (CANDELS) and the Taiwan
    ECDFS Near-Infrared Survey (TENIS) in addition to other
    data. We also revisit multi-wavelength counterparts for
    published X-ray sources from the 4Ms-CDFS and 250ks-ECDFS
    surveys, finding reliable counterparts for 1207 out of 1259
    sources ($\sim$96\%).
    Data used for photometric redshifts include intermediate-band photometry
    deblended using the TFIT method, which is used for the first time in this work. Photometric redshifts for X-ray
    source counterparts are based on a new library of AGN/galaxy
    hybrid templates appropriate for the faint X-ray population in
    the CDFS. Photometric redshift accuracy for
    normal galaxies is  0.010 and for X-ray sources is 0.014, and
    outlier fractions are 4\% and 5.4\% respectively. The results
    within the CANDELS coverage area are even better as demonstrated both
    by spectroscopic comparison and by galaxy-pair statistics.
    Intermediate-band photometry, even if shallow, is valuable when
    combined with deep broad-band photometry. For best accuracy,
    templates must include emission lines.

\end{abstract}


\section{Introduction}
 For correctly modeling galaxy evolution, the availability of
  accurate redshifts for both normal galaxies and active galactic
  nuclei (AGN) is crucial. 
Although redshifts measured via spectroscopic observations are very
reliable, they are time consuming. Long exposure times are required
for the faint sources typically found in deep field observations, and
the relatively low sky density of AGN means that it is difficult to
obtain large samples. Furthermore, spectroscopic
observations have observational limits such as the redshift range
available to optical spectrographs and the telluric
OH lines for observations with near-infrared (NIR) spectrographs
from the ground. This restricts the availability of spectroscopic
redshifts (spec-z), in particular for deep pencil-beam surveys. About
65\% of sources in the Cosmic Assembly Near-IR Deep Legacy Survey
\citep[CANDELS;][]{gro11,koe11} in the GOODS-S region are fainter
than $H = 25$, beyond any reasonable spectroscopic limit. Similarly,
only about 60\% of the X-ray sources in the 4~Ms Chandra Deep
Field-South (4Ms-CDFS) survey have reliable spec-z
\citep{xue11}. Therefore, a large number of accurate photometric
redshifts (photo-z) are needed, particularly at the faint and high-redshift
ends of the source distribution.

 For normal galaxies, previous work has achieved photo-z 
  accuracy (defined as $1.48\times \mathrm{median} (\frac {
    |\Delta z |}{ 1+z_{{s}}})$) of $\sim 0.01$ by using
  well-verified spectral energy distribution (SED) templates for
  galaxies in many fields \citep{ilb09,car10}. Within the
  Extended Chandra Deep Field-South (ECDFS), photo-z for many samples
  are available in the literature (e.g.,
  \citealt{zhe04,gra06,wuy08,car10,luo10,dah13} ). 
Although the accuracy reported in each paper is similar, 
discrepancies emerge when comparing photo-z for objects without
spectroscopic information, especially at high redshift and for faint
sources. Deep NIR observations are necessary to obtain reliable redshifts at $z> 1.5$, 
where the prominent 4000~{\AA} break shifts to NIR wavelengths.

Photo-z accuracy also depends on
the number and resolution of wavelength bands available
as already shown by \citet{ben09}. 
One of the fields with the greatest
number of photometric bands is the Great Observatories Origins Deep
Survey-South \citep[GOODS-S;][]{gia04}, which has been observed
recurrently as new facilities have become available. The GOODS-S
region is at the moment in a unique niche as homogeneous and deep
data (including the exquisite X-ray coverage with {\it Chandra}) are
available. In addition to 
intermediate-band photometry from {\it Subaru} \citep{car10} and deep
{\it Spitzer}/IRAC data \citep{dam11,ash13},  {\it HST}/WFC3 NIR data from the
CANDELS survey and $J$ and $K_{S}$ bands from the Taiwan ECDFS
Near-Infrared Survey \citep[TENIS;][]{hsi12} are now available. 
The availability of these new data will improve  the already high
accuracy of photo-z for galaxies.

Even with the best data, 
photometric redshifts for AGNs remain challenging
\citep{sal09, sal11}. Photo-z errors for AGNs can have a significant impact on galaxy/AGN
coevolution studies. For example, \citet{ros13a} found that at high
redshifts the AGNs tend to have bluer colors than inactive galaxies,
implying younger stellar populations and higher specific star
formation rates in the AGN hosts. This result, as \citeauthor{ros13a} mentioned, may
be biased by the spectroscopic selection effect and  photo-z errors
leading to a bluer host color. Low accuracy of AGN  photo-z also
affects study of the evolution of the X-ray luminosity
function (XLF) of AGNs. \citet{air10} argued that 
luminosity-dependent density evolution with a flattening faint-end
slope of the XLF at $z \geq 1.2$ may result from
catastrophic photo-z failures caused by observational
limitations and  improper templates used for photo-z computation.
For all these reasons, it is important to understand how to improve
AGN photo-z accuracy, especially for the faintest and highest-redshift AGNs.

 The situation for AGNs is further complicated by the need
  for an association with multi-wavelength data
  before photo-z can be calculated. 
  This makes the accuracy of positions for the X-ray sources and the
method and data used for the associations of crucial importance. 
Uncertain positions or different depths and wavelengths covered by
the data may yield different counterparts, and often multiple
potential counterparts exist. A simple match in
coordinates often fails to yield a reliable counterpart to any
given X-ray source. Several works have instead used the likelihood ratio method (e.g.,
\citealt{sut92, bru07, lai09,luo10, xue11,civ12}), which relies on
homogeneous coverage in a given visible or infrared
band. Most works repeat the association for several different
reference bands and finally choose a counterpart by
comparing the results. 

The past decade has witnessed important developments
in normal galaxy photo-z both by SED fitting and by machine
learning techniques, and some of these improvements can be directly
used for AGNs. For example, improvement of template-fitting
photo-z by adding emission lines to the 
templates has been demonstrated by \citet{gab04,gab06} and
\citet{ilb06}. Intermediate- and narrow-band (IB, NB) photometry is valuable to pinpoint
emission lines in the SEDs \citep{ilb09,sal09,car10,mat12}, as
simulations \citep[e.g.,][]{ben09} have predicted. Additional
improvements for AGNs are to take variability and X-ray intensity 
related to optical/infrared emission into account \citep{sal09, sal11}.
   
 The main goal of this paper is to release homogeneously computed
  photo-z for both normal galaxies and X-ray-detected AGNs in the GOODS-S,
  CDFS, and ECDFS and to provide a new X-ray source list compiled
  from the literature along with new optical/NIR/MIR associations.
The paper is organized as follows: 
Section~\ref{datasets} introduces the photometric and
spectroscopic data sets used for photo-z computation and analysis.
Section~\ref{match} associates X-ray sources with
optical/NIR/MIR counterparts using a new Bayesian 
method. Two different X-ray catalogs for the CDFS field are
available, and we discuss the differences and the implications for the
association of the counterparts. Section~\ref{galphz} presents
the photo-z results for normal galaxies,
showing the improvement by using CANDELS photometry and 
visible-wavelength IB filters.
Section~\ref{x-phz} presents the photo-z results for X-ray sources, 
and Section~\ref{discussion} discusses key factors affecting
the photo-z results.
Section~\ref{catalog} gives details of the released catalogs,
which include redshift probability distribution functions. 
Finally, Section~\ref{summary} summarizes the work.

Throughout this paper, we adopt the AB magnitude system and assume a
cosmology with $H_{0}=70$~km~s$^{-1}$~Mpc$^{-1}$,
$\Omega_{\Lambda}=0.7$, and $\Omega _{M}=0.3$ \citep{spe03}. 

\setcounter{footnote}{0}

\section{The Data Sets } \label{datasets}

 \begin{figure}
 \centering
  \includegraphics[width=0.48\textwidth]{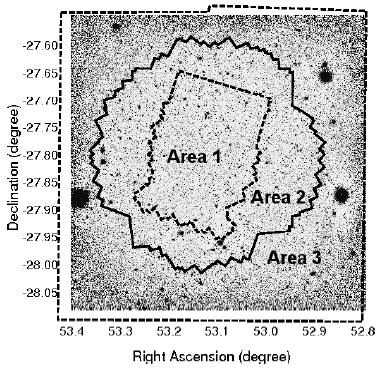}  % left, bottom, right and top
  \caption{Major areas defined in ECDFS. Background is the negative
    $J+K_s$ image from TENIS. The inner dashed line encloses  the
    CANDELS/GOODS-S area (``Area~1''), the solid line encloses the
    deep X-ray coverage (CDFS, ``Area~2''), and the outer
    dashed line (ECDFS) shows the MUSYC \citep{car10} coverage
    (``Area~3'') that defines the full area used in this paper.
      \label{zone} }
\end{figure}

The area centered on the GOODS-S field has been observed repeatedly
with a large variety of facilities and instruments. As a result,
numerous datasets with different bands and depths
are available depending on the exact location. Reliable
X-ray-to-optical associations and photometric redshifts can be
obtained only when the data are homogeneous, and for this reason we
split the area into subregions where the data are
uniform. Three main regions share the same sets of
data: Area~1 ($\sim$176~arcmin$^2$) is the region covered by CANDELS and
GOODS-S, Area~2 ($\sim$ 290~arcmin$^2$) is the 
outer CDFS region surrounding CANDELS/GOODS-S, and Area~3
($\sim$ 435~arcmin$^2$) is the ECDFS
region outside the CDFS. Figure~\ref{zone} shows the three regions.

\subsection{Photometric data from UV to MIR} \label{optdata}
Altogether the ECDFS has been covered by 50 bands from ultraviolet (UV) 
to mid-infrared (MIR) as
listed in Table~\ref{photometry}. Table~\ref{sets} summarizes the
catalogs used in each area.

\begin{table*}\footnotesize
\begin{center}
\caption{Photometric Data\label{photometry}}
\begin{tabular}{lllllc} 
\toprule[1.5pt]
Filter  & $\lambda_{\mathrm{eff}}$ & FWHM& $5\sigma$ Limiting Depth& Instrument &Area\\
&(\AA{})&(\AA{}) &(AB mag) &Telescope& \\
\midrule
$U$-CTIO\tablenotemark{a}  &3734 &387 &26.63 &Blanco/Mosaic-II &1 \\
$U$-VIMOS\tablenotemark{a} &3722 &297&27.97& VLT/VIMOS  &1\\
F435W\tablenotemark{a} &4317 &920 &28.95/30.55\tablenotemark{f}&HST/ACS &1\\
F606W\tablenotemark{a} &5918 &2324&29.35/31.05\tablenotemark{f}&HST/ACS &1\\
F775W\tablenotemark{a} &7693&1511&28.55/30.85\tablenotemark{f}&HST/ACS &1\\
F814W\tablenotemark{a} &8047&1826&28.84&HST/ACS &1\\
F850LP\tablenotemark{a}&9055&1236&28.55/30.25\tablenotemark{f}&HST/ACS &1\\
F098M\tablenotemark{a} &9851&1696&28.77&HST/WFC3 &1\\ 
F105W\tablenotemark{a} &10550&2916&27.45/28.45/29.45\tablenotemark{g}&HST/WFC3 &1\\
F125W\tablenotemark{a} &12486&3005&27.66/28.34/29.78\tablenotemark{g}&HST/WFC3 &1\\
F140W\tablenotemark{a} &13635&3947&26.89/29.84\tablenotemark{h}&HST/WFC3 &1\\
F160W\tablenotemark{a} &15370&2874&27.36/28.16/29.74\tablenotemark{g}&HST/WFC3 &1\\
$Ks$-ISAAC\tablenotemark{a} &21605&2746&25.09& VLT/ISAAC &1\\
$Ks$-HAWKI\tablenotemark{a} &21463&3250&26.45&  VLT/HAWK-I &1\\
$3.6 \mu \mathrm{m}$-SEDS\tablenotemark{a} &35508&7432&26.52& Spitzer/IRAC &1\\
$4.5 \mu \mathrm{m}$-SEDS\tablenotemark{a} &44960&10097&26.25& Spitzer/IRAC&1\\ 
$5.8 \mu \mathrm{m}$-GOODS\tablenotemark{a} &57245&13912&23.75& Spitzer/IRAC&1\\
$8.0 \mu \mathrm{m}$-GOODS\tablenotemark{a} &78840&28312&23.72& Spitzer/IRAC&1\\
$3.6 \mu \mathrm{m}$-SIMPLE\tablenotemark{b} &35508&7432&23.89& Spitzer/IRAC &2, 3\\
$4.5 \mu \mathrm{m}$-SIMPLE\tablenotemark{b} &44960&10097&23.75& Spitzer/IRAC& 2, 3\\ 
$5.8 \mu \mathrm{m}$-SIMPLE\tablenotemark{b} &57245&13912&22.42& Spitzer/IRAC& 2, 3\\
$8.0 \mu \mathrm{m}$-SIMPLE\tablenotemark{b} &78840&28312&22.50& Spitzer/IRAC& 2, 3\\
$U38$\tablenotemark{b} &3706 &357&25.33&WFI/ESO MPG &2, 3\\
$U$\tablenotemark{b} &3528 &625&25.86& ESO MPG/WFI&2, 3\\
$B$\tablenotemark{b} &4554 &915&26.45& ESO MPG/WFI&2, 3\\
$V$\tablenotemark{b} &5343 &900&26.27& ESO MPG/WFI&2, 3\\
$R$\tablenotemark{b} &6411 &1602&26.37& ESO MPG/WFI&2, 3\\
$I$\tablenotemark{b} &8554 &1504&24.30& ESO MPG/WF&2, 3\\
$z$\tablenotemark{b} &8989 &1285&23.69& Blanco/Mosaic-II&2, 3\\
$J$\tablenotemark{b} &12395 &1620&22.44& Blanco/ISPI&2, 3\\
$H$\tablenotemark{b} &16154 &2950&22.46& ESO NTT/SofI&2, 3\\
$K$\tablenotemark{b} &21142 &3312&21.98& Blanco/ISPI&2, 3\\
$J$\tablenotemark{c}  &12481 &1588&24.50&CFHT/WIRCam&2, 3 \\
$Ks$\tablenotemark{c}  &21338 &3270&23.90&CFHT/WIRCam&2, 3\\
FUV\tablenotemark{e} &1543 &228 &25.69 &GALEX &1, 2, 3\\ 
NUV\tablenotemark{e} &2278 &796 &25.99 &GALEX &1, 2, 3\\	
IA427\tablenotemark{b,d} &4253 &210&25.01&Subaru&1, 2, 3\\
IA445\tablenotemark{b,d} &4445 &204&25.18&Subaru&1, 2, 3\\
IA464\tablenotemark{b,d} &4631 &216&24.38&Subaru&1, 2, 3\\
IA484\tablenotemark{b,d} &4843 &230&26.22&Subaru&1, 2, 3\\
IA505\tablenotemark{b,d} &5059 &234&25.29&Subaru&1, 2, 3\\ 
IA527\tablenotemark{b,d} &5256 &243&26.18&Subaru&1, 2, 3\\
IA550\tablenotemark{b,d} &5492 &276&25.45&Subaru&1, 2, 3\\
IA574\tablenotemark{b,d} &5760 &276&25.16&Subaru&1, 2, 3\\
IA598\tablenotemark{b,d} &6003 &297&26.05&Subaru&1, 2, 3\\
IA624\tablenotemark{b,d} &6227 &300&25.91&Subaru&1, 2, 3\\
IA651\tablenotemark{b,d} &6491 &324&26.14&Subaru&1, 2, 3\\
IA679\tablenotemark{b,d} &6778 &339&26.02&Subaru&1, 2, 3\\
IA709\tablenotemark{b,d} &7070 &321&24.52&Subaru&1, 2, 3\\
IA738\tablenotemark{b,d} &7356 &324&25.93&Subaru&1, 2, 3\\
IA768\tablenotemark{b,d} &7676 &366&24.92&Subaru&1, 2, 3\\
IA797\tablenotemark{b,d} &7962 &354&24.69&Subaru&1, 2, 3\\
IA827\tablenotemark{b,d} &8243 &339&23.60&Subaru&1, 2, 3\\
IA856\tablenotemark{b,d} &8562 &324&24.41&Subaru&1, 2, 3\\
\bottomrule[1.5pt]
\end{tabular}
\tablenotetext{1}{CANDELS-TFIT catalog \citep{guo13}}
\tablenotetext{2}{MUSYC catalog \citep{car10}}
\tablenotetext{3}{TENIS catalog \citep{hsi12}}
\tablenotetext{4}{IB-TFIT catalog (Donley et~al.\ in preparation)} 
\tablenotetext{5}{GALEX DR6/7}
\tablenotetext{6}{Measurements from two regions: GOODS-S and
  HUDF09. See the detail in \citet{guo13} } 
\tablenotetext{7}{Measurements from three regions: CANDELS wide,
  deep, and HUDF09. See  \citet{guo13} for details.} 
\tablenotetext{8}{Measurements from two regions: 3D-HST and
  HUDF12. This is an updated version of \citet{guo13}} 
\end{center}
\end{table*}

{\bf 1. Area 1}: In this region, we primarily used the CANDELS-TFIT
multi-wavelength catalog of \citet[][hereafter G13]{guo13}, which
covers the CANDELS GOODS-S area with 18 broad-band filters mostly 
from space observatories. The photometry was based on
template-fitting \citep[TFIT;][]{lai07}, using the high resolution
WFC3/$H$-band image to detect sources and define apertures, which
were then used for photometry in lower-resolution
images. TFIT was also applied to the MIR data from the
Spitzer Extended Deep Survey \citep[SEDS;][]{ash13}. This 
deblending yields more accurate photo-z
and also increases the probability of making correct X-ray to IR
associations (see Section~\ref{match}). In addition, the Area~1 data
include 18 IBs at optical wavelengths provided by
the MUSYC team\footnote{Multi-wavelength Survey by Yale-Chile. 
The reduced images are available at \url{http://www.astro.yale.edu/MUSYC/} }
\citep{car10}. CANDELS collaborators (Donley et~al.\ in
preparation) have produced an IB-TFIT catalog with the same
parameters used by G13. Despite being up
to 2 magnitudes shallower than the rest of the optical data, the IB
data are useful to identify emission lines, which
can modify the choice of template best fitting the data and thus
the photo-z (see Section~\ref{impact_em}). To these 36 bands we also
added the near-UV (NUV) and far-UV (FUV) data from GALEX Data Release
6 and 7. 
The association between GALEX data and the WFC3/$H$-band catalog was
done via positional matching within a radius of 1\arcsec. About 5\% 
of all sources and $\sim$25\% of X-ray-detected sources
have UV counterparts. The combined data, which we refer to as
``\tf '', have 34930 sources with up to 38 bands for computing photo-z.\\

{\bf 2. Areas 2+3}: These areas differ in depth of X-ray coverage
(Section~\ref{xdata}) but have the same data sets otherwise. For the
CDFS and ECDFS surrounding Area~1, 
we merged the following photometric catalogs via coordinate cross
match, allowing a maximum separation of 1\arcsec: (1) GALEX catalog (as
above), (2) the original MUSYC catalog \citep{car10}, and (3) the
$J$ and $K_s$-band data from the Taiwan ECDFS Near-Infrared
Survey\footnote{The TENIS data are available at
  \url{http://www.asiaa.sinica.edu.tw/~bchsieh/TENIS/About.html} }
\citep[TENIS;][]{hsi12}. Although TENIS
is no deeper than existing NIR data, the TENIS data are more
homogeneous over the entire field and have slightly different
transmission curves, increasing the wavelength coverage. The MIR
data for Area~2 and~3 came from the
Spitzer IRAC/MUSYC Public Legacy in ECDFS
\citep[SIMPLE;][]{dam11}. These data are shallower than the SEDS
data available in Area~1. Table~\ref{photometry} lists the data sets
used, and we refer to this dataset as ``MUSYC+TENIS''. There are
70049 sources in this photometry.

\subsection{X-ray data}\label{xdata} 

The X-ray catalogs to cross-match were obtained from the
{\it Chandra} survey of 4Ms-CDFS observations covering Area~1+2 
and from the 250ks-ECDFS observations covering Area~3.
Two independent groups \citep{xue11, ran13} have provided source 
catalogs for 4Ms-CDFS using different methods for data reduction and source detection. 
 Similarly for Area~3, both \citet{leh05} and \citet{vir06} have released X-ray source
  catalogs for the 250ks-ECDFS survey. We have cross-matched X-ray sources 
  from both catalogs in each area.\\

\noindent
{\bf For Areas 1+2 we used}:

\hangindent=0.9cm 
a. The 4Ms-CDFS source catalog of \cite{xue11} (hereafter X11)
with 740 point-like  X-ray sources. The sensitivity
limits of the X-ray data are $3.2\times 10^{-17}$, $9.1\times
10^{-18}$, and $5.5\times 10^{-17}$~erg~cm$^{-2}$~s$^{-1}$ for
the full (0.5--8~keV), soft (0.5--2~keV), and hard (2--8~keV) bands,
respectively. 

\hangindent=0.9cm 
b. The 4Ms-CDFS source catalog of \cite{ran13} (hereafter
R13)\footnote{ The 4Ms-CDFS X-ray catalog of R13 is available under [Surveys] $>$ [CDFS] through the portal 
    \url{http://www.mpe.mpg.de/XraySurveys}.} produced
using the analysis methodology of \citet{lai09}. The catalog contains
569  point-like X-ray  sources and has sensitivity limits
$4.2\times 10^{-17}$, $1.2\times 10^{-17}$, and $8.8\times
10^{-17}$~erg~cm$^{-2}$~s$^{-1}$ in the full, soft, and hard
bands, respectively.\\

\noindent
{\bf For Area 3 we used}:

\hangindent=0.9cm 
c. The 250ks-ECDFS X-ray catalog from \citet{leh05} (hereafter L05)
with 762 sources in the entire ECDFS of which 457 are in Area~3
(i.e., outside the 4Ms-CDFS area). Catalog 
sensitivity limits are $1.1\times
10^{-16}$~erg~cm$^{-2}$~s$^{-1}$ in the soft (0.5--2~keV) band
and $6.7\times 10^{-16}$ in the hard (2--8~keV) band. 

\hangindent=0.9cm 
 d. The 250ks-ECDFS X-ray catalog from \citet{vir06} (hereafter
  V06) with 651 sources in the entire ECDFS of which 404 are in
  Area~3. Sensitivity limits are $1.7\times
  10^{-16}$ and $3.9\times 10^{-16}$~erg~cm$^{-2}$~s$^{-1}$ in the soft and hard bands, respectively.

\subsection{Spectroscopic Data} 

\begin{figure}
  \centering
  \includegraphics[width=0.47\textwidth]{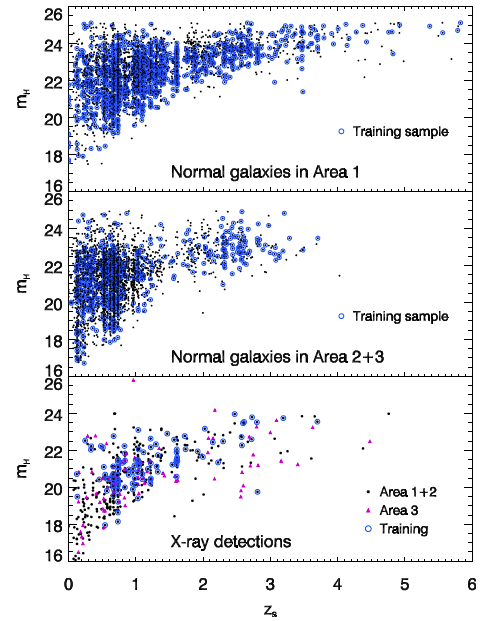}
 \caption{$H$-band magnitude as a function of  spec-z for all objects
   with spectroscopic redshifts. Black dots
   in the top  panel represent 
   normal galaxies in Area~1, where \tf\ data are available. The
   middle panel shows normal galaxies identified from the MUSYC
   catalog in Areas~2 and~3. Black dots in the bottom panel
   represent X-ray-detected sources in Areas~1 and~2, and magenta
   triangles denote sources detected in the shallower X-ray data in
   Area~3. Open blue circles in all three panels 
   indicate objects used for training.
    \label{hmag_spz} } 
\end{figure}

The availability of spec-z for a subgroup of sources is essential for
computing reliable photo-z via SED fitting \citep{dah13}. 
A subset of spec-z can first be used for training under the assumption that
they are representative of the entire population. A different subset
can then be used for testing photo-z quality. 
For this work we cross matched the photometric
catalogs to a compilation of spec-z (N.~Hathi, private
communication), allowing a maximum separation of 1\arcsec. There are
2314 ($\sim$7\%) Area~1 sources that have
reliable spec-z and 3880 ($\sim6\%$) such sources in Areas~2 and~3
(2016 in Area~2, 1864 in Area~3). 

As discussed by \citet{dah13}, optimal results are obtained when the
templates used for the photo-z computation are calibrated on the
photometry available for the spectroscopic training samples. For this
reason the training samples should fully span the entire
magnitude--redshift parameter space. Figure~\ref{hmag_spz} shows that
the 1000 sources randomly selected as our training samples
are indeed spread over all redshift and magnitude
ranges in the respective Areas. Because the
photometry available in Area~1 differs from that in Area~2+3,
two sets of training samples and computations of the zero-point
offsets\footnote{Zero-point offset is the average for each
  photometric band of the difference between the photometry of
  training set objects and photometry predicted by the best-fit
  template at the object's redshift. The offset in each
  band depends on the set of templates used  and the
  number of bands available.} were used.

For the X-ray sources, we forgo using the training sample for
computing zero-point offsets and instead use it to sample the AGN
population and help build the AGN-galaxy hybrid templates needed for
proper SED fitting and photo-z measurement
\citep{sal09}. For this purpose, we randomly chose $\sim$25\% of
the 4Ms-CDFS detections with available spec-z over the entire range
of redshift and magnitude that have CANDELS data and used
them as the
training set to build hybrid templates. The remaining $\sim$75\%
were used for unbiased testing of the results. Details are given in the
Appendix.

\begin{table*}\footnotesize
\centering
\caption{Catalogs used for redshift estimation and counterpart identification.\label{sets}}
\begin{tabular}{llll}
\toprule[1.5pt]
  &Area 1 &Area 2 &Area 3  \\ 
\midrule
 &4Ms-CDFS-X11\tablenotemark{e}       &4Ms-CDFS-X11        &250ks-ECDFS-L05\tablenotemark{i}   \\
 &4Ms-CDFS-R13\tablenotemark{f}        &4Ms-CDFS-R13        &250ks-ECDFS-V06\tablenotemark{j}   \\
Cross           &CANDELS-TFIT                                  &MUSYC                    &MUSYC             \\
matching      &MUSYC                                             &TENIS                       &TENIS    \\ 
                     &TENIS                                               &SIMPLE-IRAC\tablenotemark{h} & SIMPLE-IRAC\\   
                     &SEDS-IRAC\tablenotemark{g}  & &\\
\midrule
\multirow{3}{*}{Photo-z}
&CANDELS-TFIT\tablenotemark{a}     &MUSYC\tablenotemark{c}          &MUSYC          \\	
&IB-TFIT\tablenotemark{b}    &TENIS\tablenotemark{d}  &TENIS  \\
&GALEX-UV        &GALEX-UV      &GALEX-UV       \\
\midrule
$N_\mathrm{spz}$ & 2314&2016&1864\\
\bottomrule[1.5pt]
\end{tabular}
\tablecomments{$N_\mathrm{spz}$ is the number of spec-z used in each Area (N.~Hathi, private communication)} 
\footnotetext[1]{\citet{guo13}} 
\footnotetext[2]{Donley et~al.\ in preparation}
\footnotetext[3]{\citet{car10}}
\footnotetext[4]{\citet{hsi12}}
\footnotetext[5]{\citet{xue11}}
\footnotetext[6]{\citet{ran13}}
\footnotetext[7]{\citet{ash13}}
\footnotetext[8]{\citet{dam11}}
\footnotetext[9]{\citet{leh05}}
\footnotetext[9]{\citet{vir06}}
\end{table*}

\vspace{10mm}

\section{X-ray to optical/NIR/MIR Associations}\label{match}
 
 X-ray source positions can differ between catalogs because of
  different methods adopted for data reduction and source detection.
The goal of this paper is not to judge which method of X-ray source
detection is superior but rather to provide accurate photo-z for
optical/NIR/MIR sources associated with
X-ray sources. Associations between X-ray sources and possible
  counterparts were therefore done independently for each of the four
  X-ray catalogs (Sec~\ref{xdata}), and duplicate sources were
  removed only at the end of the process as described below.

\subsection{Comparing X-ray Catalogs} \label{x11r13}

\noindent
{\bf 1. In Areas 1+2:} 

 The major difference between the R13 and X11 catalogs is that
  R13 adopted a higher threshold for source detection.
  Despite that, there are some sources in the R13 catalog 
  but not in X11.  There are also astrometric differences, which
can affect the association to an optical/NIR/MIR
counterpart. Thus the redshift assigned to the X-ray source
and also to the supposed counterparts can be different because 
different template libraries and priors were used for X-ray
galaxies than for normal ones.  In order to match X-ray catalogs,
we shifted the X11 positions by  $-0\farcs175$ in
R.A.\ and $0\farcs284$ in Dec.\footnote{The original X11 positions
  are on the radio astrometric frame. The shifts needed to bring them to
  the optical frame are in Sec.~3.1 of the X11 paper.} 
to register them to the optical frame \citep{gia04}. The
R13 catalog is already on the MUSYC optical frame and was not shifted.

\begin{figure}[h]
  \centering
  \includegraphics[width=0.45\textwidth]{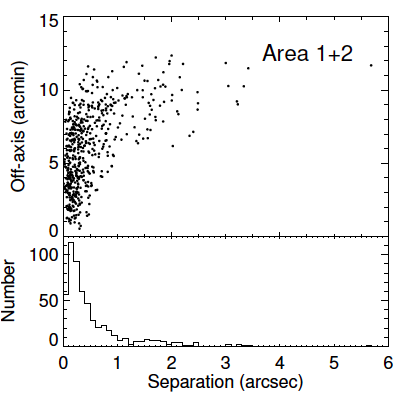}
  \caption{Coordinate differences between X11 and R13 X-ray catalogs.
    The lower panel shows a histogram of offsets for the 545 sources
    Area~1 and~2 in common in the two catalogs. The upper panel
    shows the off-axis angle from the {\it Chandra} aim point as a
    function of the angular offset.
  \label{offset1}}
\end{figure}

  After astrometric shifting, we matched the X11 and
  R13 catalog coordinates, allowing a maximum distance of
  10\arcsec. There are 545 sources in common with a
  maximum offset $<6\arcsec$ as shown in
  Figure~\ref{offset1}. For these 545 sources, neither catalog has
  any additional X-ray source within 10\arcsec. As expected, all of
  the large offsets are for sources at large off-axis angles. For off-axis
  angles $<6\arcmin$, the median coordinate offset is
  0\farcs13, and except for one source, the maximum
  offset at any off-axis angle is $<3\farcs5$.
  We treat each of the 545 matched sources as a single X-ray
  detection. $54\%$ of these sources have a distance from 
  each other larger than the positional error claimed for either of the catalogs. 
  In addition, there are 195 sources
  detected by X11 but not R13 and 24 sources
  detected by R13 but not X11 for a total of
  764 X-ray sources in Areas 1+2.  As R13 mentioned, the unique sources to 
  either of the two catalogs are mostly low-significance
  detections and therefore of lower reliability. In the following discussions,  
  ``X-'' sources indicate those from X11 and ``R-'' those from R13. \\

\noindent
{\bf 2. In Area 3: }

\begin{figure}[h]
  \centering
  \includegraphics[width=0.45\textwidth]{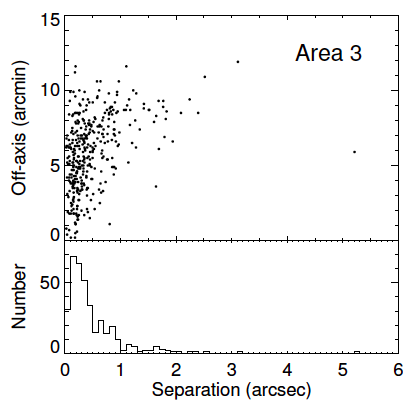}
  \caption{Coordinate differences between L05 and V06 X-ray catalogs.
    The lower panel shows a histogram of offsets for the 495 sources
    in Area~3 in common in the two catalogs. The upper panel shows
    the off-axis angle from the {\it Chandra} aim point as a function
    of the angular offset.
   \label{offset2}}
\end{figure}

We adopted the
  \citet{car08} astrometric calibration to align the V06
  positions to the MUSYC and
  L05 catalogs, which were already in agreement. After the shift, the
  two catalogs 
  have 366 source matches with offsets $<6\arcsec$. These have
  a median separation of 0\farcs16 (Fig.~\ref{offset2}). We
  consider these 366 sources as the same X-ray detection. $12\%$ of 
  these sources have a separation that is larger than the positional 
  error associated with either of the catalogs.
  In addition, there are 91 sources in the L05 catalog but not
  V06 and 38 sources in the V06 catalog but not L05 for a total of
  495 X-ray sources in Area~3. 
  A compilation of the four X-ray catalogs with their original positions and fluxes is available under [Surveys] $>$ [CDFS] through the portal 
    \url{http://www.mpe.mpg.de/XraySurveys}.

\subsection{Matching method}\label{sec:matching_method}

We used a new association method based on Bayesian statistics which
allows pairing of sources from more than two catalogs at once while
also making use of priors. Salvato et~al.\ (2014, in preparation)
will give details, but in brief the code finds matches based on
the equations developed by \citet{bud08}. Then additional probability
terms based on the magnitude and color distributions are applied.
(See \citealt{nay13} for a similar approach.) The code was
developed in view of the launch of eROSITA \citep{mer12}, where an
expected million sources will be scattered over the entire sky and
will have a non-negligible positional error and/or non-homogenous
multi-wavelength coverage, conditions not optimal for association
methods like Maximum Likelihood (e.g., \citealt{bru07, luo10,
  civ12}). The new method provides the same quality of results 
  as Maximum Likelihood in a much shorter time because
matches are done simultaneously across all bands. Thus for example
sources that are extremely faint or undetected in optical bands but
brighter in the IRAC 3.6~$\mu \mathrm{m}$ band can be
identified as counterparts in a single iteration.

  For the 4Ms-CDFS sources (X11 and R13) located in Area 1, we used
  the CANDELS/$H$-selected catalog,
  TENIS/\linebreak[0]$J\&K_s$-selected catalog, 
  MUSYC/\linebreak[0]$BVR$-selected catalog, and the deblended
  SEDS/IRAC 3.6~\micron\
  catalog. For the 4Ms-CDFS sources located in Area~2,
  we matched the X-ray sources to the TENIS/$J\&K_s$-selected catalog,
  MUSYC/$BVR$-selected catalog, and SIMPLE/IRAC 3.6$\mu \mathrm{m}$
  catalog. The same set of these three catalogs was also used in Area~3 
  for finding the associations for the 250ks-ECDFS sources (L05 and V06).
  Table~\ref{sets} summarizes the catalogs matched in each Area. 
  
\begin{figure}
 \centering
  \includegraphics[width=0.45\textwidth]{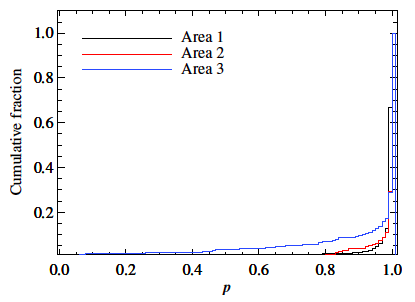}
 \caption{Cumulative fraction of the posterior $p$ for the possible counterparts to 
 the X-ray sources in Area 1, 2, and 3 as indicated in the legend. \label{post}}  
\end{figure}   

For each X-ray source (740 from X11, 569 from R13, 440 from 
L05, and 374 from V06), we considered all catalog
objects lying within 4\arcsec\ of the X-ray position and computed the
posterior probability $p$ that the given object
is the correct counterpart. 
Figure~\ref{post} shows the distribution of the posteriors for all
the possible associations in the three areas. 
 In Area 1 where the data are deeper and better resolved, more than $98\%$ of the
 X-ray sources have at least one association with $p>0.7$, and we consider this $p$ value the threshold for 
defining an association in all three areas. Area 3 has a distribution of $p$ that reaches lower values, 
but because of the shallowness and lower resolution of the data, we consider not reliable the association with $p<0.7$. 
Our catalogs (see Sec.~\ref{catalog}) include the
$p$ value to allow users to define a stricter threshold,
depending on the scientific use intended.

\begin{figure}[h]
   \epsscale{1.15}
   \plotone{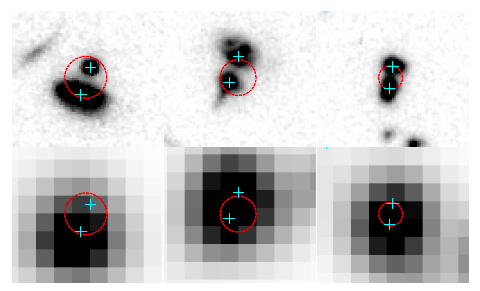}    
   \caption{Three examples of multiple $H$-band associations (from
     left to right X-115, X-517, and X-224) in $H$-band (upper) and
     IRAC-3.6$\mu \mathrm{m}$ (lower). The size of each cutout is
     $5'' \times 5''$ . The red circles are centered at the X-ray
     position with radius corresponding to the positional error. The
     cyan crosses indicate the positions of $H$-band detected sources
     from G13. These three cases have two $H$-band associations both
     with probabilities greater then 0.99. The uses of deblended
     IRAC photometry does not help in making a unique secure
     association.\label{close}}
\end{figure}

Figure~\ref{close} shows examples of ambiguous identifications. In all
three cases shown, a single X-ray source has two possible 
$H$-band associations with $p>0.99$. Even the
simultaneous use of deblended IRAC photometry from SEDS does not
help in associating a unique counterpart. The example in the middle shows
that despite the high resolution of the CANDELS images,
the upper source is still blended, and
probably a third component is present. If a further deblending were
applied, the $H$ flux  would be split among multiple
components, thus reducing the probability of the upper source being
the right association. In practice, we attempted no further
deblending and simply
flagged these kinds of objects as sources with multiple counterparts. For
these cases, in addition to the photo-z computed using normal galaxy
templates, we also provide the values obtained by assuming that they
are AGNs. The photo-z results reveal that $\sim 20\%$ of these
  close pairs have similar redshifts and may be associated with
  galaxy mergers or galaxy groups. However the
  majority of apparent pairs are projections of unrelated objects.

\subsection{Matching results} \label{match_result} 

\begin{figure}
   \centering
   \includegraphics [width=0.47\textwidth]{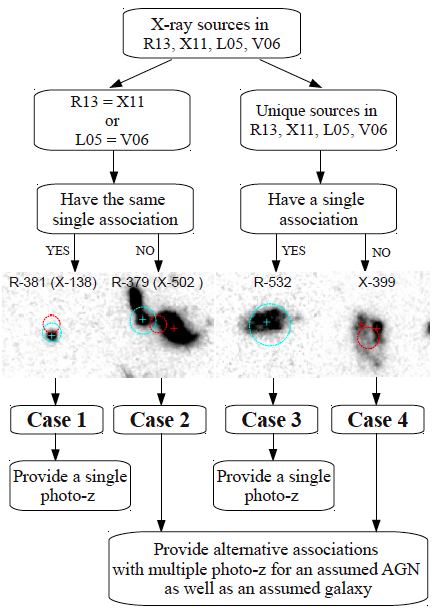}
   \caption{Flow chart of the process for four cases of
     X-ray to optical/NIR/MIR associations. $H$-band negative images ($5'' \times 5''$) are
     provided as examples for each case . Dashed-line
     circles show the X11 (red) and R13 (cyan) X-ray
     positions and  positional uncertainties. Red and
     cyan crosses show the corresponding $H$-band counterparts.
     \label{flowchart}}
\end{figure}

Figure~\ref{flowchart} shows the decision tree for X-ray source
associations and computing  photo-z, and Table~\ref{xcase} gives
numbers for each case in each Area. There are four cases:\\
   
\hangindent=0.9cm 
{\bf 1. Case~1:} An X-ray source in both catalogs with one
optical/NIR/MIR association. Case~1 means the same unique
  association was chosen even though the X-ray catalog positions may differ
  between X11 and R13 or between L05 and V06. There are
  714 of these sources  in Areas~1+2+3.

\hangindent=0.9cm 
{\bf 2. Case~2:} An X-ray source in both catalogs
  with differing optical/NIR/MIR  associations. Case~2 can arise from
  two causes:  (1) position differences in the  X-ray catalogs may point to different
  counterparts, or (2) there may be more than one potential counterpart near
  the X-ray position(s), and we
  cannot tell which is the right one. Some of the latter may be blended sources
  with more than one galaxy contributing to the X-ray flux.
  In total, there are 181 case~2 sources
  in Areas~1+2+3. These sources are
  identified in the final catalogs, and counterpart photo-z are
  calculated using both AGN and normal galaxy SED templates (see
  Section~\ref{catalog}) . 

\hangindent=0.9cm 
{\bf 3. Case~3:} X-ray sources found in one catalog but not the other, having
  a unique counterpart. There are
  235 of these sources in Areas~1+2+3. 

\hangindent=0.9cm 
{\bf 4. Case~4:} X-ray sources found in only one catalog and having
  multiple possible counterparts. There
  are 77 such sources in Areas~1+2+3. As for
  Case~2, the catalogs identify all the possible counterparts and provide
  both AGN and normal galaxy photo-z results for each.\\

 In summary, 1207 out of 1259 ($\sim$96\%) of the X-ray
  sources are associated with multi-wavelength counterparts, and 258
  of them ($\sim$21\%) have multiple counterparts possible. 
  There are 26 sources for which the counterpart is detected only in the
IRAC bands, and no photo-z computation is possible for these. All the
other sources have at least six photometric points, and a photo-z is
provided. The photo-z catalog (see Sec.~\ref{catalog}) entry for each
source indicates the number of photometric points used for the
photo-z computation. The remaining 52 sources ($\sim 4\%$) either 
have no identifications in any of the optical/NIR/MIR catalogs ($\sim 1\%$) or 
have possible counterparts identified with $p<0.7$ ($\sim 3\%$). 
For these sources, the photo-z are not available as well. 

\begin{table*}
\centering
\caption{Results of X-ray to optical/NIR/MIR associations in ECDFS . \label{xcase}}
%\begin{tabular}{cc|cccc|cccccc}
\begin{tabular}{cccccccccccc}
\toprule[1.5pt]
&$N_{x}$ & Case1& Case2& Case3& Case4&$N_{ctp}^{single}$&$N_{ctp}^{multi}$ &$N_{ctp}$& $N_{ctp}^{multi}/N_{ctp} $ & $N_{ctp}/N_{x} $  \\ 
\midrule
Area 1    &509  &272  &67  &130  &29  &402 &96  &498    &19\%        &$98\%$ \\
\midrule
Area 2    &255  &170  &29  &35    &12    &205  &41  &246   &17\%       &$96\%$ \\
 \midrule
Area 3    &495  &272  &85  &70   &36  &342  &121  &463    &26\%       &$94\%$ \\
 \midrule
TOTAL   &1259 &714  &181  &235  &77 &949  &258  &1207  &21\%   &$96\%$ \\
\bottomrule[1.5pt]
\end{tabular}
\tablecomments{$N_{x}$: Number of X-ray sources; $N_{ctp}^{single}$:
  Number of sources that have only one possible counterpart;
  $N_{ctp}^{multi}$:  Number of sources that have more than one
  possible counterpart; $N_{ctp}$: Total number of sources for which
  at least a counterpart was found. } 
\end{table*}

\begin{figure*}
  \centering
   \includegraphics[width=0.5\textwidth]{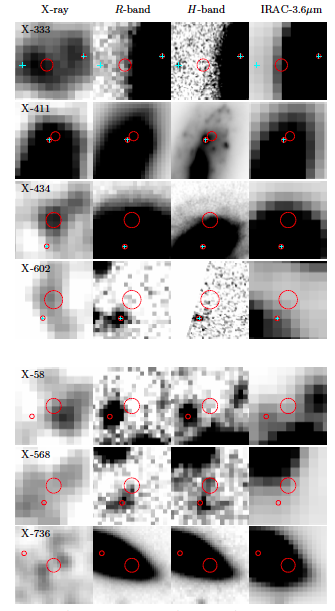}
  \caption{Multi-wavelength images of the seven sources from X11 for
    which we found new, secure ($p>0.9$) counterparts. Wavelengths
    are indicated above each set of panels. The four
    sources in the upper group are in Area~1 and have CANDELS
    $H$-band images. The three sources in the lower group have no
    WFC3-$H$, and TENIS-$K_s$  is shown instead. X-ray images are
    full-band from X11. The red dashed circles are centered at the X11
    positions with their radii showing the corresponding positional
    uncertainty. Cyan crosses in the upper panels show all $H$-band
    detections, and the solid red circles show the catalog position
    of the chosen counterpart. All cutouts are  $5'' \times 5''$
    except that X-736 is $10'' \times
    10''$. \label{not_onir}} 
\end{figure*}

\subsection{Comparison to previous results in Area 1+2}\label{sec:comp_match}

X11 used  likelihood ratio matching to assign counterparts to 716 out
of 740 X-ray  sources in Areas~1+2. 
  Our code and the newly available ancillary data give
  secure counterparts (with $p>0.9$) for seven additional sources
  shown in Figure~\ref{not_onir}. Most of the new counterparts
  are offset from the X-ray position by 1--2 times the X-ray position uncertainty.
  The most likely reason for finding new identifications is having
  better imaging data available, but there remains a chance some of the
  X-ray sources are not real.

\begin{figure}
  \centering
  \includegraphics[width=0.48\textwidth]{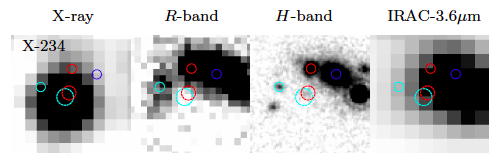}
   \caption{ Negative images of the source
     R-57 (=X-234). Image wavelengths are indicated at the top, and
     each image is $5'' \times 5''$. Red dashed-line circles are centered at the position
     provided by X11 and cyan dashed-line circles at the position
     given by R13.
     Circle sizes indicate the respective X-ray position uncertainties. Red and cyan solid-line
     circles are the counterparts we assign to the two X-ray positions, and
     the blue circle indicates the counterpart assigned by
     X11. \label{234} } 
\end{figure}

  Figure~\ref{234} shows an example of a revised X-ray
  association. In this case low resolution catalogs give a single
  counterpart for the source (R-57=X-234) for either X-ray position 
  However, the high-resolution WFC3/$H$-band image
  reveals at least four sources close together,
  and the slightly different coordinates provided by X11 and R13
  point to different but equally likely counterparts.
 This difference is mainly due to the
  catalogs chosen for cross matching rather than the
  matching method. The Bayesian method should in principle give 
  the same result as the maximum likelihood method, but the
  ability to match several catalogs simultaneously greatly 
  improves the efficiency of the matching.

\section{Photo-z for Non X-ray detected Galaxies}\label{galphz}

  This section focusses on the X-ray-undetected
  sources, which we refer to as ``normal galaxies'' even though some
  will in fact be AGNs.\footnote{A large fraction of
  galaxies that host  AGNs in their central regions do not emit detectable
  X-rays but are identified at infrared and/or radio 
  wavelengths or by emission line ratios \citep[e.g.,][]{don12}. } 
 The derived photo-z will be reliable to the extent of the ``normal galaxies'' which have normal
  galaxy SEDs at observed visible/infrared wavelengths.
  X-ray sources need more tuning for accurate photo-z and are
  discussed in Section~\ref{x-phz}.

The photo-z were computed using {\it LePhare} \citep{arn99,ilb06}, which
is based on a $\chi^{2}$ template-fitting method.
For the normal galaxies, we adopted the same templates,
extinction laws, and absolute magnitude priors as 
\citet{ilb09}. In short, 31 stellar population templates were corrected for 
theoretical emission lines by modeling the fluxes with line ratios of
[\ion{O}{3}]/[\ion{O}{2}], H$\beta$/[\ion{O}{2}],
H$\alpha$/[\ion{O}{2}], and Ly$\alpha$/[\ion{O}{2}].
In addition to the galaxy
templates, we also included a complete library of star templates as did 
\citet{ilb09} and \citet{sal09}.
Four extinction laws (those of \citealt{pre84}, \citealt{cal00}, and two
modifications of the latter, depending on the kind of templates) were
used with $E(B-V)$ values of 0.00 to 0.50 in steps of 0.05~mag.
Photo-z values were allowed to reach $z=7$ (in steps of 0.01) because
deeper photometry allows us to reach higher redshifts (see details
given by \citealt{ilb09}). The fitting procedure included a
magnitude prior, forcing sources to have an absolute magnitude in
rest $B$-band between $-8$ and $-24$. Photometric zero-point corrections
were incorporated but never exceeded 0.1~mag. Final best parameters
came from minimizing $\chi^2$. We advise against this step for the AGNs because optical variability is intrinsic to the source and not accounted for in the photometry.

 All the normal galaxies were selected from either
  the CANDELS-TFIT catalog or the MUSYC catalog (Sec.~\ref{optdata}). Photo-z are 
  based on \tf\ photometry for sources detected in the CANDELS-TFIT
  catalog and otherwise on MUSYC+TENIS photometry. The majority of
  normal galaxies have \tf\ photometry in Area~1 
  but only MUSCY+TENIS photometry in Areas~2+3. \\

Quantifying the photo-z
accuracy ($\sigma_{\mathrm{NMAD}}$)\footnote{Our measure of photo-z   
  accuracy is the normalized median absolute deviation (NMAD):
  $\sigma_{\mathrm{NMAD}} \equiv 1.48\times \mathrm{median} (\frac {
    |\Delta z |}{ 1+z_{{s}}}$), where $z_{{s}}$ is
  spec-z, $z_{{p}}$ is photo-z, and $\Delta
  z\equiv(z_{{p}}-z_{{s}})$. Outliers were not removed before
  computing $\sigma_{\mathrm{NMAD}}$.},
the percentage of the outliers ($\eta$)\footnote{Outliers are defined as
  $\frac{ | \Delta z|}{1+z_{{s}}} > 0.15$},  and the mean offset
between photo-z and spec-z
($\mathrm{bias}_{z})$\footnote{$\mathrm{bias}_{z} \equiv 
  \mathrm{mean} (\frac { \Delta z}{1+z_{{s}}})$ after
  excluding outliers.} was based on the spectroscopic
samples. Table~\ref{gal_phz_tab} gives these measures
of photo-z quality for the global samples and for
subsamples split into magnitude and redshift bins.

\subsection{Area 1}\label{galphz_zone1}

\begin{figure}
  \centering
  \includegraphics[width=0.47\textwidth]{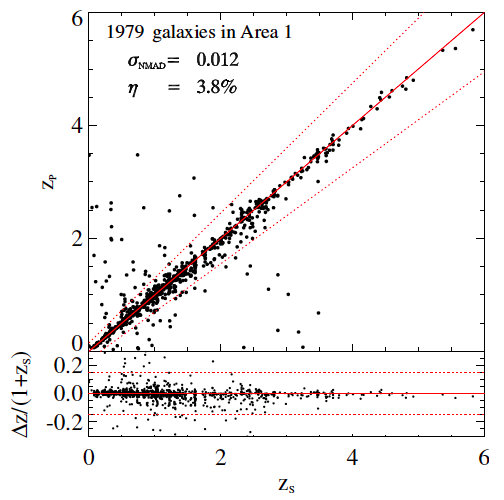}
   \caption{Upper panel: Photo-z vs spec-z. Dots represent all normal
     galaxies with spec-z  in Area~1. The solid 
     line represents $z_{{p}}= z_{{s}}$; the dotted
     lines represent $z_{{p}}= z_{{s}} \pm 0.15(1+
     z_{{s}})$. Lower panel: Same but plotted as the
     difference $\Delta z\equiv(z_{{p}}-z_{{s}})$. \label{gal_phz1}}
\end{figure}

The overall outlier fraction of $\sim$3.8\% in this region
is comparable to the most recent work by the CANDELS team
\citep{dah13}. However, the
deblended IB photometry from MUSYC improves the accuracy to
$\sigma_{\mathrm{NMAD}} =0.012$ (from $\sigma_{\mathrm{NMAD}} =0.026$
by \citealt{dah13}) and $\mathrm{bias}_{z} =-0.001$ (from
$\mathrm{bias}_{z} =-0.005$).
Figure~\ref{gal_phz1}
illustrates the results. Outlier fractions and
scatter are larger for the fainter galaxies
(Table~\ref{gal_phz_tab}), but bias is only a weak function of source
magnitude. Bias is, however, larger for $z>1.5$ galaxies than for
those at lower redshifts. Scatter and outlier fraction are also
larger at $z>1.5$, but this mostly reflects the typically fainter magnitudes of
the more distant sources.

\begin{figure}
 \centering
  \includegraphics[width=0.45\textwidth]{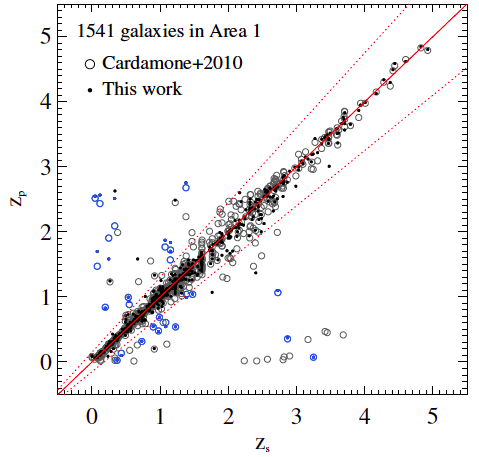}
 \caption{Photo-z vs. spec-z for 1541 normal galaxies in Area~1. Black dots 
 are results from this work, and grey open circles are results from \citet{car10}. 
 Blue dots and blue open circles indicate objects that are outliers both in our 
 work and in that of \citet{car10}. \label{gs_gal_car} \\}  
\end{figure}

The decreased outlier fraction in the present survey requires {\it both} the 
deeper CANDELS-TFIT data and the deblended IB photometry.
Table~\ref{CvsM} gives data quality measures for 1541 sources in
common using various data sets. Using only MUSYC+TENIS, but not the deep \tf\ data, produces the same data quality as
\citet{car10} as expected.
However using the \tf\
photometry decreases the outlier fraction from $\sim$4\% to
$\sim$2\%, and the decrease is most substantial (more than a factor
of two) for the faint and distant
sources. (See Table~\ref{CvsM}.) 
Figure~\ref{gs_gal_car} illustrates the comparison and in particular
the decrease in outliers at $z_s>2$. The difference comes from the
use of deep space-based data (i.e., CANDELS) and the TFIT technique
for deblending the lower-resolution bands. However, the IB data are
also important. \citet{dah13} used the CANDELS-TFIT data, while
their results (included in Table~\ref{CvsM}) are better than with the
ground-based data alone, they 
are not as good as with the combined data sets (i.e., \tf ). Adding the IB data improves results---mainly in
accuracy but also in outlier fraction---even for the fainter subset of the sample.

\begin{table*}
\begin{center}
\caption{Photo-z Quality  for Normal Galaxies\label{gal_phz_tab}}
\begin{tabular}{ccccccccccccc}
\toprule[1.5pt]
  \multicolumn{1}{c}{}&\multicolumn{4}{c}{Area 1}&\multicolumn{4}{c}{Area 2+3} &\multicolumn{4}{c}{Area 1+2+3} \\
  \cmidrule(r){2-5} \cmidrule(r){6-9} \cmidrule(r){10-13} 
 & $N$ &$\mathrm{bias}_{z}$ & $\sigma_{\mathrm{NMAD}}$ & $\eta(\%)$  & $N$&$\mathrm{bias}_{z}$& $\sigma_{\mathrm{NMAD}}$& $\eta(\%)$ & $N$&$bias_{z}$& $\sigma_{\mathrm{NMAD}}$& $\eta(\%)$ \\
\midrule
    Total       &1979   &-0.001   &0.012    &3.79    &3444    &0.001   &0.009    &4.21  &5423      &0.001      &0.010        &4.06\\
\midrule
 $R < 23$    &576    &0.003   &0.008    &1.04   &2414    &0.001   &0.009    &2.20   &2990      &0.001      &0.009        &1.97\\
 $R > 23$   &1403   &-0.002   &0.015    &4.92   &1030    &0.002   &0.012    &8.93  &2433     &-0.001      &0.013        &6.62\\
 \midrule
 $H < 23$   &1323   &-0.000   &0.011    &2.87   &2428    &0.002   &0.009    &2.72  &3751      &0.001      &0.009        &2.77\\
 $H > 23$    &656   &-0.002   &0.016    &5.64   &1016   &-0.001   &0.011    &7.78   &1672     &-0.001      &0.012        &6.9\\
 \midrule
 $z < 1.5$   &1652   & 0.002  &0.011    &3.51   &3316    &0.002   &0.009    &3.89   &4968      &0.002      &0.009        &3.76\\
 $z > 1.5$    &327   &-0.013  &0.021    &5.20   & 128   &-0.008   &0.031   &12.50    &455     &-0.011      &0.024        &7.25\\
\bottomrule[1.5pt]
\end{tabular}
\end{center}
\end{table*}

\begin{table*}
\begin{center}
\caption{Comparison of Photo-z Results for Normal Galaxies in Area~1\label{CvsM}}
\begin{tabular}{cccccccccccccc}
\toprule[1.5pt]
  \multicolumn{2}{c}{}&\multicolumn{3}{c}{$\rm{TFIT}_{\rm{CANDELS+IB}}$}&\multicolumn{3}{c}{MUSYC+TENIS}&\multicolumn{3}{c}{Cardamone+2010} &\multicolumn{3}{c}{Dahlen+2013} \\
  \cmidrule(r){3-5} \cmidrule(r){6-8} \cmidrule(r){9-11} \cmidrule(r){12-14} 
 & $N$ &$\mathrm{bias}_{z}$ & $\sigma_{\mathrm{NMAD}}$ & $\eta(\%)$ &$\mathrm{bias}_{z}$& $\sigma_{\mathrm{NMAD}}$& $\eta(\%)$&$\mathrm{bias}_{z}$& $\sigma_{\mathrm{NMAD}}$& $\eta(\%)$&$\mathrm{bias}_{z}$& $\sigma_{\mathrm{NMAD}}$& $\eta(\%)$ \\
\midrule
    Total   &1541   &0.000   &0.011   & 2.14   &0.003   &0.012    &3.96 & 0.000   &0.011   & 3.96  &-0.005   &0.026    &2.47  \\
\hline
 $R < 23$    &506    &0.003   &0.009    &0.79    &0.002   &0.008    &0.79   &0.002   &0.008 &0.99  &-0.002   &0.026  &0.99\\
 $R > 23$   &1035   &-0.002   &0.013    &2.80   &0.003   &0.016    &5.51   &-0.001   &0.016 &5.41 &-0.006   &0.026  &3.19\\
 $H < 23$   &1064   & 0.001   &0.010   & 1.60    &0.003   &0.010    &2.07  & 0.000   &0.010   & 2.07 &-0.006   &0.027 &1.97\\
 $H > 23$   & 477   &-0.002   &0.014   & 3.35    &0.002   &0.021    &8.18   & 0.000   &0.022   & 8.18 &-0.001   &0.024  &3.56\\
 $z < 1.5$   &1308    &0.002   &0.010    &2.14    &0.004   &0.011    &3.13  &0.002   &0.010 &2.98  &-0.005   &0.026  &2.45\\
 $z > 1.5$   & 233   &-0.014   &0.019    &2.15   &-0.002   &0.030    &8.58  &-0.008   &0.045 &9.44 &-0.002   &0.023  &2.58\\
\bottomrule[1.5pt]
\end{tabular}
\end{center}
\end{table*}

\subsection{Areas 2 and 3} \label{galphz_zone23}

\begin{figure}
  \centering
  \includegraphics[width=0.47\textwidth]{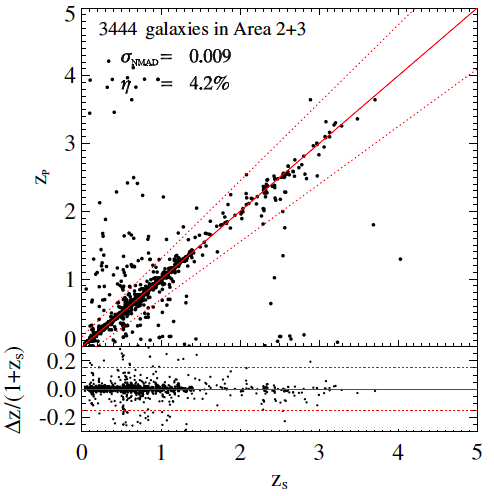}
 \caption{Photo-z vs spec-z of normal galaxies in Areas~2+3. }. \label{gal_phz23}
\end{figure}

Outside the CANDELS area, photo-z quality using MUSYC+TENIS
photometry is similar to that of \citet{car10}.
Figure~\ref{gal_phz23} illustrates the results. The brighter and
lower-redshift subsets have photo-z quality almost as good as in
Area~1 (see Table~\ref{gal_phz_tab}), but fainter galaxies have a
higher outlier fraction. This is just as expected from the tests in
Section~\ref{galphz_zone1}.\\

\begin{figure}
  \centering
  \includegraphics[width=0.47\textwidth]{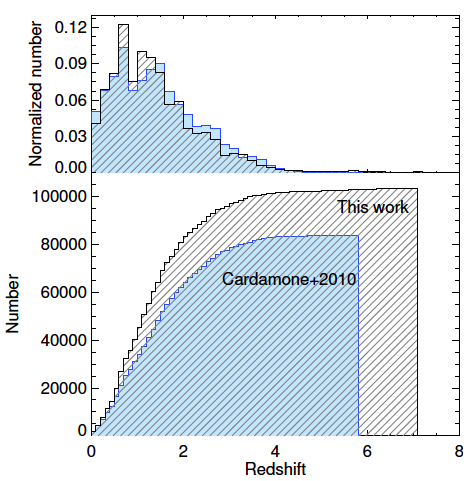}
  \caption{Upper panel: normalized photo-z distribution for 
    normal galaxies. Grey hatched area shows results of this work,
    and blue shaded area shows results of \citet{car10}/  Lower
    panel: cumulative number of 
    normal galaxy photo-z redshifts for this work and for
    \citet{car10} as labeled. \label{gal_phz_hist}} 
\end{figure}

The entire ECDFS (Areas~1+2+3) contains $\sim104000$ normal galaxies that have
photo-z up to  $z\sim7$. 
Figure~\ref{gal_phz_hist} shows the advantage of using WFC3 NIR to
detect more sources in total and especially at $z\ga2$.
An interesting paradox is that we actually 
have a slightly lower fraction of sources at $z>1.5$ than \citet{car10}.
This is probably because  their higher outlier fraction, lacking
deep NIR data, leads to more outliers with apparent  $z>1.5$. 

\section{Photo-z for X-ray sources }\label{x-phz}

\begin{figure*}
  \centering
        \includegraphics [width=1.0\textwidth]{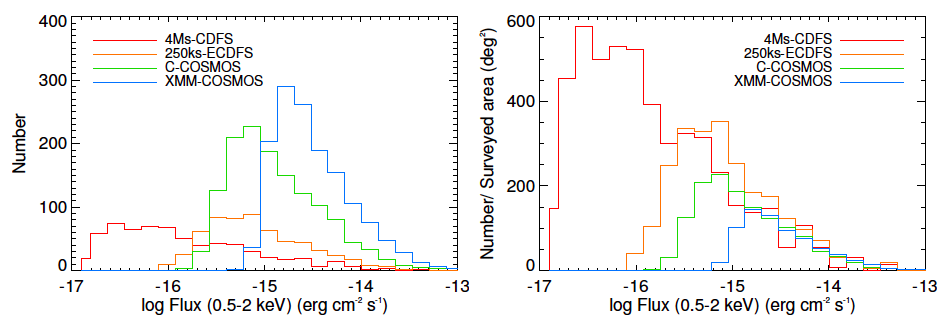}
   \caption{Soft X-ray flux distributions in numbers (left) and
     source densities (right). Histograms show distributions for
     the 4Ms-CDFS (Areas~1 and ~2), 250ks-ECDFS (Area~3), and
     comparison surveys 
       {\it Chandra}-COSMOS \citep{elv09} and {\it XMM}-COSMOS
       \citep{cap09}  surveys as indicated in the legend. \label{flux}} 
\end{figure*} 

 AGNs require special treatment to calculate photo-z. This paper
  uses deep X-ray data to identify candidate AGNs. However, 
  X-ray surveys as deep as the 4Ms-CDFS also detect significant
  numbers of star-forming galaxies. The library must therefore
  include templates of normal galaxies, AGNs, and hybrids.

\subsection{Template Library Methods}\label{tlm}

 \citet{luo10}
  computed photo-z for sources in the 2Ms-CDFS
  by using the spectra of known sources as templates for  SED
  fitting. Using the entire spectroscopic sample for
  template training gave an apparent accuracy
  $\sim$0.01 with almost no outliers. However, unbiased testing
  suggested a true accuracy of 0.059 and $\sim$9\% outliers. This demonstrates
  how important the training sample is. For the
same field, \citet{car10} created hybrid templates by combining normal galaxy templates with
the SED of a type~1 AGN. The method gave accurate results ($\sigma_{\rm NMAD}\sim
0.01$) but large outlier fraction ($\sim 12\%$). \citet[][hereafter S09, S11]{sal09,sal11}  pursued a
different approach for X-ray sources in the 
COSMOS field \citep{sco07} detected by {\it XMM} \citep{cap09} and by {\it
  Chandra} \citep{elv09}. 
This involved (1) correcting the photometry for variability
when applicable; (2) separating the optical counterparts to the X-ray
sources into two subgroups: point-like and/or variable sources in one
and extended, constant sources in the other; (3) applying absolute magnitude priors 
to these two subgroups, assuming that the former are AGN-dominated
while the latter are galaxy-dominated; (4) creating AGN-galaxy
hybrids, using different libraries for the two subgroups. This same
procedure substantially reduced the fraction of outliers and gave
higher accuracy than standard photo-z techniques when
applied to X-ray sources in COSMOS. The procedure has also yielded
reliable results for 
the Lockman Hole \citep{fot12} and AEGIS fields (Nandra et al.\ 2014,
submitted).
S11 also verified the need for depth-dependent template libraries by
showing that hybrids defined for {\it XMM}-COSMOS are
not optimal for the deeper
{\it Chandra}-COSMOS. Even though the X-ray-faint {\it Chandra} 
sources are AGNs (i.e., $L_{\rm{x}}
>10^{42}$), normal galaxy templates  gave better results for them
than AGN-dominated templates.

\subsection{Constructing Population-dependent SED Libraries}  \label{xpop}

 For this work, we constructed new hybrid templates following the procedure of 
S09 and S11 (Sec.~\ref{tlm}). First we point out that the difference of the X-ray flux distributions between the two X-ray surveys used in this work (i.e., 4Ms-CDFS and 250ks-ECDFS) is even more extreme than what we have found in S11 (i.e., XMM-COSMOS and Chandra-COSMOS). 

Fig.~\ref{flux} shows the soft X-ray flux distributions of the 4Ms and 250ks sources, together with the distributions from the Chandra-COSMOS \citep{elv09} and XMM-COSMOS \citep{cap09}. 
The left panel shows the distribution in numbers for each survey. Because of the sky coverage and the depth of the observations, most of the 4Ms-CDFS sources are located in the faint part of the flux distribution, which is opposite to the locus occupied by the shallower observations (e.g. XMM-/Chandra-COSMOS and 250ks-ECDFS). After normalizing by the total surveyed area\footnote{This is not the Log N-Log S} (see the right panel), it reveals that the X-ray bright sources which are similar to the XMM-COSMOS sources are very rare in the 4Ms survey. This implies that the library of hybrids used in XMM-COSMOS is probably not representative of the 4Ms population. Based on this considerations, we need to build a new library for the fainter 4Ms-CDFS population.
 The Appendix gives details, but in short AGN SEDs were combined
in various proportions with semi-empirical galaxy SEDs (the same as
already successfully used by \citet{gab04}, \citet{dro05}, and
\citet{feu05}) from the FORS Deep Field \citep{ben01} to make hybrid templates. 
The AGN SEDs were modified QSO1 and QSO2 originally from
\citet{pol07}. Separate libraries were used for (a) point-like
sources in Areas~1+2, (b) point-like sources in Area~3, and (c) extended
sources in all Areas. Because the flux distribution of point-like
sources in Area~3 is similar to that of the {\it XMM}-COSMOS field,
library (b) for point-like sources in that area was the same as used
by \citet{sal09}. 

As a first step, we split the sources into extended and point-like subgroups 
depending on the observed source FWHM
in the WFC3/$H$-band images for Area~1 and
the MUSYC/$BVR$ images for Areas~2 and~3. The
  extended sources were assumed to be host-dominated, and being seen
  as extended means they are likely to be at low redshift.
  For these sources, we applied an absolute
  magnitude prior $-24<M_{B}<-8$ and used
  templates with at most a small AGN fraction. Point-like sources are usually
  AGN-dominated and can be at any redshift. We therefore applied a prior
  $-30<M_{B}<-20$ to these and used hybrid AGN-galaxy templates. 
  The library of stellar templates was the same as 
  used by \citet{ilb09} and \citet{sal09}.

 For the {\it XMM}-COSMOS field, \citet{sal09} had multi-wavelength,
  multi-epoch observations spanning several years. About 1/4 of
  sources seen in those were variable. The lack of multi-epoch
  data for the CDFS/ECDFS means that we cannot detect the variable objects
  and correct their 
  photometry. However, these objects are a minor contributor to the
  X-ray population in the much smaller CDFS
  area (1/15 of XMM-COSMOS area).
 Therefore only a minor fraction of the Area~1 
  and~2 sources are likely to be variable.
  The major effect of being unable to correct for variability will
  be an increased outlier fraction rather 
  than decreased photo-z accuracy
  \citep{sal09}. Area~3, covered at 250~ks depth, is an intermediate
  case, and part of the photo-z inaccuracy there could be due to lack of
  variability correction. The spectroscopic testing in the respective Areas
  (Table~\ref{xmix_phz_tab}) quantifies the outlier fractions
  and the inaccuracies resulting from all causes.

\begin{table*}\footnotesize
\begin{center}
\caption{Photo-z Quality for X-ray Sources\label{xmix_phz_tab}}
\begin{tabular}{ccccccccccccccccc}
\toprule[1.5pt]
  \multicolumn{1}{c}{}&\multicolumn{4}{c}{Area 1}&\multicolumn{4}{c}{Area 2}&\multicolumn{4}{c}{Area 3}&\multicolumn{4}{c}{Area 1+2+3} \\
  \cmidrule(r){2-5} \cmidrule(r){6-9} \cmidrule(r){10-13}\cmidrule(r){14-17}
& $N$  &$bias_{z}$&$\sigma_{\mathrm{NMAD}}$ & $\eta(\%)$ & $N$
  &$\mathrm{bias}_{z}$& $\sigma_{\mathrm{NMAD}}$& $\eta(\%)$ & $N$
  &$\mathrm{bias}_{z}$&$\sigma_{\mathrm{NMAD}}$ & $\eta(\%)$&$N$
  &$\mathrm{bias}_{z}$&$\sigma_{\mathrm{NMAD}}$ & $\eta(\%)$ \\ 
\midrule
Total      & 300      &-0.002       &0.012        & 2.67     & 104      &-0.002       &0.014        & 6.73    & 148      &-0.004       &0.016        &10.14    & 552      &-0.002       &0.014        & 5.43  \\
\hline
   $R < 23$    & 171      &-0.004       &0.010        & 1.17    &  80      &-0.000       &0.014        & 5.00    & 109      & 0.001       &0.013        & 8.26   & 360      &-0.002       &0.011        & 4.17\\
   $R > 23$    & 129      & 0.001       &0.024        & 4.65    &  24      &-0.008       &0.016        &12.50    &  39      &-0.018       &0.023        &15.38   & 192      &-0.004       &0.023        & 7.81\\
   $H < 23$    & 278      &-0.003       &0.012        & 1.80    & 102      &-0.002       &0.014        & 6.86    &  69      &-0.004       &0.016        &10.14   & 449      &-0.003       &0.013        & 4.23\\
   $H > 23$    &  22      & 0.012       &0.014        &13.64    &   2      &-0.010       &0.026        & 0.00   &  79      &-0.004       &0.016        &10.13    & 103      &-0.001       &0.014        &10.68\\
  $z < 1.5$    & 240      &-0.001       &0.012        & 2.50    &  86      &-0.001       &0.015        & 4.65    & 112      &-0.003       &0.020        & 8.04   & 438      &-0.001       &0.014        & 4.34\\
  $z > 1.5$   &  60      &-0.004       &0.014        & 3.33    &  18      &-0.009       &0.012        &16.67    &  36      &-0.009       &0.010        &16.67   & 114      &-0.006       &0.012        & 9.65\\
%\hline
%  \multicolumn{1}{c}{Library}&\multicolumn{4}{c}{Lib-EXT/PT}&\multicolumn{4}{c}{Lib-EXT/PT}&\multicolumn{4}{c}{Lib-EXT + S09} \\
\bottomrule[1.5pt]
\end{tabular}
%\tablecomments{Lib-EXT/PT are libraries for extended and point-like source built in this work. S09 represents the libraries of hybrids from \citet{sal09}.}
\end{center}
\end{table*}

\subsection{Results}\label{xresult}

\begin{figure}
\includegraphics [width=0.45\textwidth]{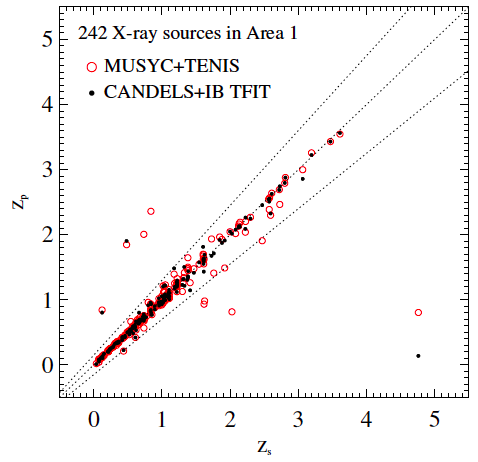}
\caption{Comparison of photo-z to spec-z with and without
  \tf\ photometry. Filled points show results for 242 X-ray sources
  from the \citet{car10} catalog using the full \tf\ dataset. Open
  circles show results for the same sources using only the
  MUSYC+TENIS data.
  \label{TFIT_vs_MUSYC_xmix}}
\end{figure}

\begin{figure}[htb]
\includegraphics [width=0.45\textwidth]{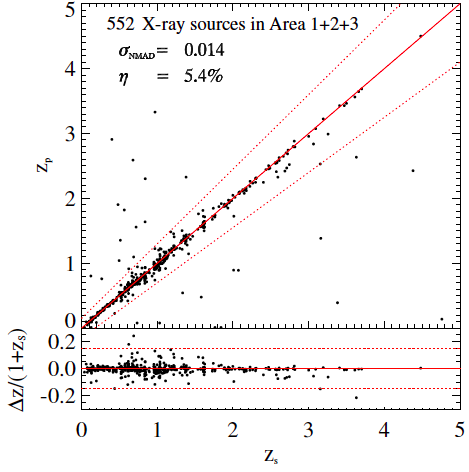}
  \caption{Comparison of photo-z to spec-z for all X-ray sources in Areas
    1+2+3.\label{zone123_phz_fig}} 
\end{figure}

In Area~1, where \tf\ photometry and high-resolution space-based
images are available, the photo-z for X-ray
sources (Table~\ref{xmix_phz_tab}) are as accurate as those for
normal galaxies (Table~\ref{gal_phz_tab}). Remarkably, the outlier
fraction is actually lower for the X-ray sources than for normal
galaxies. Excellent photo-z quality is 
maintained even for $z>1.5$.
Figure~\ref{TFIT_vs_MUSYC_xmix} shows that the results are largely
attributable to the WFC3 data with their high angular resolution.
Instead of using ground-based data (i.e., MUSYC+TENIS), the use of $\rm{TFIT}_{\rm{CANDELS+IB}}$ catalog allows us to reduce the outlier fraction by a factor of 3. 
The improvement is especially
great for the $R>23$ and $z>1.5$ sources. The outlier fractions decreases from 
$6.3\%$ to $2.1\%$ for faint sources and from 
$12.8\%$ to $2.6\%$ for high-redshift sources. Comparison with Area~2 also confirms the importance of the WFC3 data.
Without these data, photo-z accuracy deteriorates only slightly
(Table~\ref{xmix_phz_tab}), but the outlier fraction triples.
Most of the outlier increase comes from the $R>23$ and $z>1.5$ subsets.
(There are only two sources with $H>23$, and numerical results
for that bin are meaningless.)

 Area~3 has a larger fraction of outliers than either of the
  other two Areas, though accuracy for the non-outliers is little
  worse than in Areas~1 and~2 (Table~\ref{xmix_phz_tab}). Three
  effects probably contribute to the larger fraction of outliers. One is 
  shallower photometry at the border of the field (Fig.~\ref{zone}),
  leading to larger errors. Second, the X-ray coverage is shallower 
  in the larger Area~3, thus the fraction of varying Type~1 AGN 
  is presumably higher. The lack of 
  variability correction will therefore have a larger effect.
  This is likely exacerbated by the third effect, having to use
  ground-based images rather than higher-resolution 
  images for classifying sources as point-like or extended.
  In Area~1, about 30\% of
  sources are classified point-like using WFC3 but extended on
  a ground-based image, due to the low resolution of the images and being sensitive to the presence of nearby sources. 
  Using the template library for the extended sources rather than for point-like 
  classification would have doubled the outlier fraction.

\begin{figure}
\includegraphics [width=0.45\textwidth]{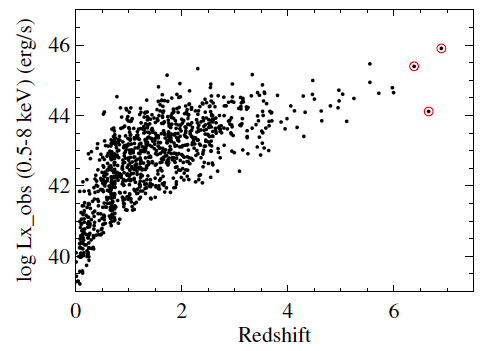}
\caption{Distribution of 0.5--8~keV observed-frame X-ray
  luminosity as a function of redshift for all X-ray
  sources. Redshifts are spec-z if available and otherwise
  photo-z. Red open circles indicate the three anomalous
  sources that have unreliable photo-z.
  }\label{lx_z}
\end{figure}

 Furthermore, in order to identify possible outliers among the sources 
without spec-z, we look at the distribution of observed-frame 
X-ray luminosity as a function of redshift. In Figure~\ref{lx_z}, three sources with apparent extreme redshift are probably
  outliers. They are located on the edge of the optical images and have unreliable
  or non-existent MUSYC photometry, leaving only six photometric data points 
  (from the TENIS catalog). Photo-z with so few data points cannot be trusted.

\section{Discussion}\label{discussion}
\subsection{Photometric Redshift Accuracy Beyond the Spectroscopic Limit}

\begin{figure*}
\centering
      \includegraphics[width=\textwidth]{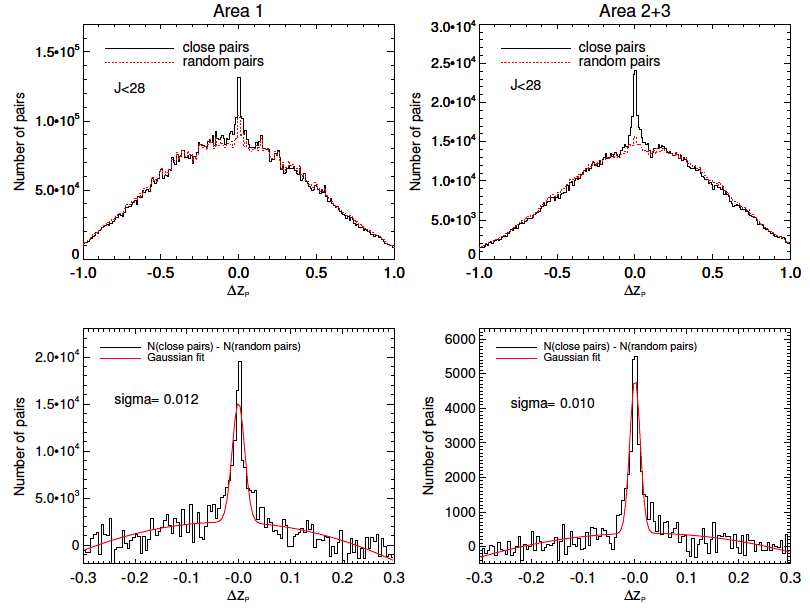}
 \caption{Distributions of photo-z differences for pairs of
   galaxies. In the upper panels, red dotted lines represent
   differences for random
   pairs and black solid lines represent differences for pairs
   having angular separation $<$15\arcsec. Only galaxies with $J<28$
   are included. The lower panels show results for
   close pairs after 
   subtracting the distributions for random pairs. Black lines show
   the observed $\Delta z_p$, and
   red lines show a Gaussian fit with standard deviation sigma as
   indicated in each panel. The left two panels are for Area~1 and
   the right two for Areas~2+3.
   \label{pair_stat}}
\end{figure*} 

Using spec-z to estimate photo-z accuracy (as in
Tables~\ref{gal_phz_tab} and~\ref{xmix_phz_tab}) 
is not representative of 
sources fainter than the spectroscopic limit. 
\citet{qua10} introduced a method for estimating photo-z accuracy 
based on the tendency of galaxies to cluster in space.
Because of clustering, galaxies seen close to each other on the
sky have a significant probability of being 
physically associated and having the same redshift.
Therefore the
distribution of photo-z differences\footnote{Photo-z difference is
  defined as $\Delta
  z_p\equiv(z_{{p,1}}-z_{{p,2}})/(1+z_{\rm{mean}}) $.}
($\Delta z_{{p}}$) of close pairs will show an excess at small
redshift differences over the distribution for random pairs. This is
seen in Figure~\ref{pair_stat}.\footnote{Random pairs also show a
  noticeable peak at small $\Delta z_{{p}}$. This is not due to any
  systematic we can identify and may be due to the known large scale
  structure \citep{cast07,sali09,deh14} in the field.} 
The excess for close pairs with
magnitude $J< 28$ fits a Gaussian with standard deviation
  $\sigma =0.012$ in Area~1 and $0.010$ in
Areas 2+3. Because the width includes the
scatter from both paired galaxies,
the photo-z uncertainty for an individual object should be $\sigma /
\sqrt{2}$. These values are given in Table~\ref{pair_tab}.\footnote{
The close pair excess includes only objects with similar photo-z, so 
outliers are excluded in calculating $\sigma$ here.}  The pair test  reveals that the faint sources
without spec-z have photo-z accuracy similar to that of sources
bright enough to have spectroscopic data.

\begin{table}[!h]
\centering
\caption{Photo-z Scatter from Pair Statistics\label{pair_tab}} 
\begin{tabular}{ccc}
\toprule[1.5pt]
&Area 1 &Areas 2+3 \\
\midrule
  $J < 25$  &0.008  & 0.007    \\
  $J < 26$  &0.009  &0.007    \\
  $J < 27$  &0.009  &0.007    \\
  $J < 28$  &0.008 &0.007   \\
\bottomrule[1.5pt]
\end{tabular}
\tablecomments{\protect\parbox[t]{2.4in}{Table values $\sigma / \sqrt{2}$
    are the estimated 
  standard deviation of a single galaxy photo-z as derived from
  galaxy pairs in each magnitude range.}}
\end{table}
\subsection {The Impact of Intermediate-band Photometry} \label{impact_ib}

\begin{figure}
    \includegraphics[width=0.45\textwidth]{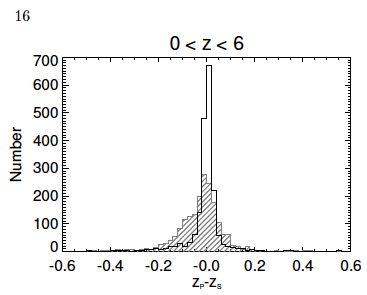}
    \caption{Distribution of photo-z minus spec-z. Histograms show
      $z_{p}-z_{s}$ distribution for all galaxies with spec-z
      in Area~1. Black line shows results for photo-z with
      IB photometry included, and hatched area shows results for
      the same galaxies with IB data omitted.
      \label{IB-all}}
\end{figure} 

\begin{figure*}
    \includegraphics[width=1\textwidth]{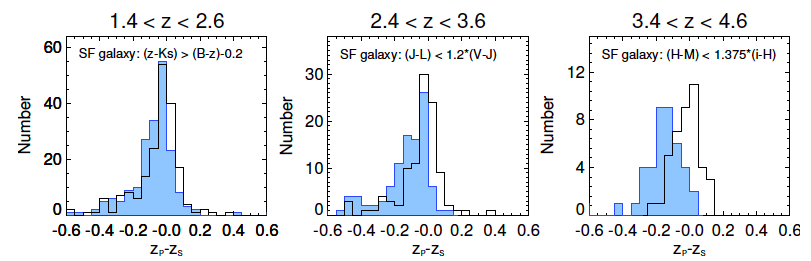}
    \caption{Distribution of photo-z minus spec-z for star-forming
      galaxies selected by rest $BzK$ colors. Histograms show
      ($z_{\rm p}-z_{\rm s}$) distributions in various redshift bins
      as indicated above each panel. Black lines show the
      distributions for photo-z with IB photometry included, and
      blue areas show distributions for the same galaxies with IB
      photometry omitted. All data are from Area~1.
       \label{IB-sb}}
\end{figure*}

Previous work has shown the importance of IB photometry for
photo-z, especially because IB data can show the presence or absence of
emission lines in galaxy SEDs. For example \citet{ilb09} showed that
including IBs improved photo-z accuracy  from 0.03 
to 0.007  for  normal
galaxies with $i^{+}<22.5$ in the COSMOS field. \citet{car10} found
the same in the ECDFS. For AGNs, \citet{sal09} 
showed that for both
extended and point-like X-ray sources in COSMOS, accuracies and
outlier fractions were substantially better when IBs were included.

In the current data, the IB photometry is much shallower than the NIR
data from CANDELS (Table~\ref{photometry}). 
To examine whether
the shallow IB data are helpful or not in this case, we recomputed
the Area~1 photo-z with exactly the 
same CANDELS-TFIT dataset \citep{guo13} used by \citet{dah13}, i.e.,
without IBs.
Results are given in Table~~\ref{CvsM}, and Figure~\ref{IB-all}
compares results with IBs and without.

Without the IBs, the outlier fraction is
5\%,  accuracy is 0.037, and $\mathrm{bias}_{z} = -0.010$. These
are similar to the
results of \citet{dah13} ``method 11H,'' which used the same
code as this 
work. The negative value of $\mathrm{bias}_{z}$ indicates 
underestimation of photo-z on average. That results in lower galaxy
luminosities and incorrect rest colors. As discussed by \citet{ros13a},
these may lead to incorrect measurements of galaxy ages and stellar
populations. The IB data
improve the accuracy and mean offset substantially, creating a
narrower and more symmetric peak of photo-z values around the spec-z
(Fig.~\ref{IB-all}).
%($\sigma_{\mathrm{NMAD}}=0.012$ and $bias_{z}=-0.001$). 

Intermediate bands should be most important for objects that have
strong emission lines in their spectra. Strong emission lines can
arise either from vigorous star formation or an AGN. To quantify the
effect, we applied (inverse) $BzK$ selection \citep{dad04} to define a sample
of star-forming objects among those with reliable spec-z. In order to
extend the selection at high redshift, we applied the revised $BzK$
criterion as defined by \citet{guo13}.\footnote{The exact criteria
  were (1) $(z-K_{s}) > (B-z)-0.2$ in the redshift range $1.4<z<2.6$;
  (2) $(J-L) > 1.2\times(V-J)$ in the redshift range $2.4<z<3.6$; (3)
  $(H-M) > 1.375\times(i-H)$ in the redshift range $3.4<z<4.6$. Symbols
  $B,\, V,\, i,\, z,\, J,\, H,\, K_{s},\, L,\, M$ refer to F435W, F606W,
  F775W, F850LP, F125W, F160W, ISAAC $K_{s}$, IRAC 3.6~$\mu
  \mathrm{m}$, IRAC 4.5~$\mu \mathrm{m}$, respectively.}
Fig.~\ref{IB-sb} shows the resulting distributions of photo-z minus
spec-z. At all redshifts, the distribution including IB is more peaked and
symmetric around zero when IBs are included.

\subsection {Impact of Emission Lines in the Templates} \label{impact_em}

\begin{figure*}
     \includegraphics[width=1\textwidth]{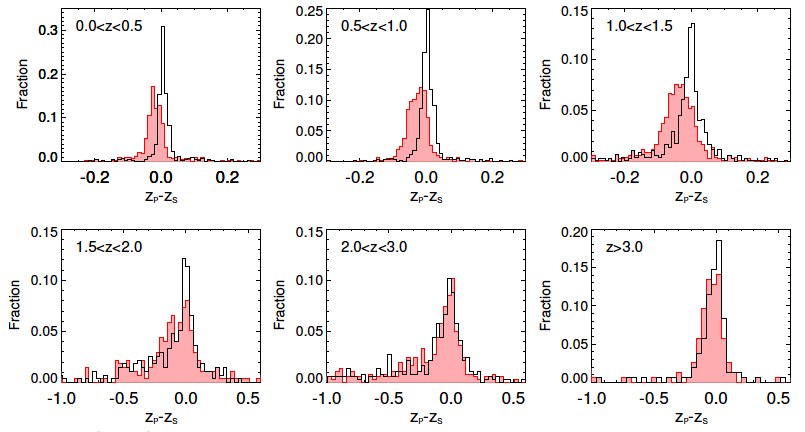} 
     \caption{($z_{\rm p}-z_{\rm s}$) distribution in various
       redshift bins. The photo-z are computed using the
       $\rm{TFIT}_{\rm{CANDELS+IB}}$ photometric catalog and using
       the templates with (black solid line ) and without (red solid
       lines) emission line contributions. The upper panel shows that
       the emission lines are useful particularly at
       $z<1.5$. \label{deltaz}}
\end{figure*} 

\citet{ilb09} demonstrated the importance of taking emission lines
into account  for
photo-z. Including lines in the templates  improved photo-z accuracy by a
factor of 2.5 for bright ($i^{+}<22.5$) galaxies in the COSMOS
field. The same effect is seen in the deeper \tf\ data as shown in
Figure~\ref{deltaz}. Although outlier numbers remain similar
($\sim$4\%) whether emission lines are included in the
templates or not, the distributions of ($z_{\rm p}-z_{\rm s}$) change.
At $z<1.5$, including emission lines gives much narrower peaks and
lower bias. At
$z>1.5$, the improvement is less than at lower
redshifts. Possible reasons are: (a) the
  contribution of the emission lines are diluted when observed in the
  NIR bands, which have broader bandwidths than optical bands; (b)
  the recipes for adding
  emission lines to the templates may be wrong for high-redshift
  galaxies; and/or (c) the IB data may be too shallow to affect 
  the high-redshift (and therefore faint)
  sources. However, even at $z>1.5$, the photo-z accuracy still shows a
  factor of 1.5 improvement ($\sigma_{\mathrm{NMAD}}$ decreasing
  from 0.032 to 0.021) when emission lines are included in the templates.

\subsection{Testing Libraries for the X-ray Population}

Because of the different X-ray populations in the 4Ms-CDFS and
250ks-ECDFS surveys, we adopted different libraries for point-like
sources in Areas 1+2 and Area~3. For
the sake of template comparison, we tried using the Area~3
(``S09'') library to calculate photo-z for point-like sources in
Areas 1+2. The fraction of outliers increased from $5.3\%$ to $15\%$,
and the accuracy became two times worse than achieved with the
preferred library. Even for $R<23$ sources, $\sigma_{\rm NMAD}$ went
from 0.011 with the proper templates to
0.016  with the old ones. For $R>23$
galaxies, the deterioration was  from 0.027 to 0.059.
In Area~3, on the other hand, using the new templates
instead of the S09 ones made photo-z slightly worse: for 
$R<23$, $\sigma_{\rm NMAD}$ was 0.009 for the new and 0.008 for the S09
libraries. For $R>23$, accuracies were 0.025 and 0.017 respectively.
Moreover $\mathrm{bias}_{z}=-0.014$ using the
S09 library but increased to $\mathrm{bias}_{z}=-0.031$ with the
new library. The better performance of the S09 library in Area~3
can be understood because the population of point-like X-ray sources
in the 250k-ECDFS is similar to the {\it XMM}-COSMOS population, and
the S09 library is more suitable for counterparts of such bright X-ray sources.

\subsection{Impact of UV data}

\begin{figure}
\includegraphics[width=0.45\textwidth]{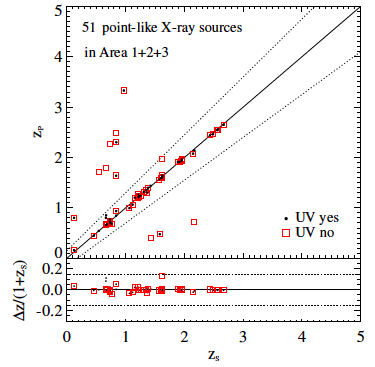}
\caption{Comparison of photo-z with spec-z for X-ray sources with
  point-like counterparts. All 51 available sources in Areas 1+2+3 are
  plotted. Black dots indicate photo-z computed with
  UV data, and red squares indicate photo-z computed without UV
  data. \label{ptuv}}
\end{figure}

UV emission from accretion disks around supermassive black
holes makes type~1 AGNs distinguishable from normal
galaxies. Therefore including UV data in the photometry is crucial
for SED fitting to obtain accurate photo-z and to decrease outliers
for AGNs. To demonstrate this we compared photo-z for AGNs
obtained with and without photometry in the UV bands. About $25\%$ of
all the X-ray detected sources in Areas~1+2+3 have UV data available
from GALEX. Among these, 221 sources have spectroscopy available and
were used as our test sample.
As expected, for the optically extended sources, where the host
dominates the emission, there is very little difference in accuracy
and fraction of outliers whether UV data are included or not.
For 170 extended sources with
spectroscopy, including UV data decreases $\sigma_{\mathrm{NMAD}}$ from
0.013 to 0.012 and $\eta$ from $5.9\%$ to $5.3\%$. In contrast, for the 51
point-like (i.e., AGN-dominated) sources, adding the
UV data halves the number of outliers (from $23.5\%$ to $11.8\%$)
though with only modest improvement in  accuracy (from 0.013 to
0.011). Among the five remaining outliers (see 
Fig.~\ref{ptuv}), two are faint ($\mathrm{mag}> 23$) in the UV, and  three
are close to other sources with the UV flux  blended in the 10\arcsec\
GALEX aperture. Deblending
the GALEX photometry with TFIT as in the other bands could perhaps
improve these cases.

\section{Released Catalogs}\label{catalog}

\begin{figure}
   \centering
      \includegraphics[width=0.45\textwidth]{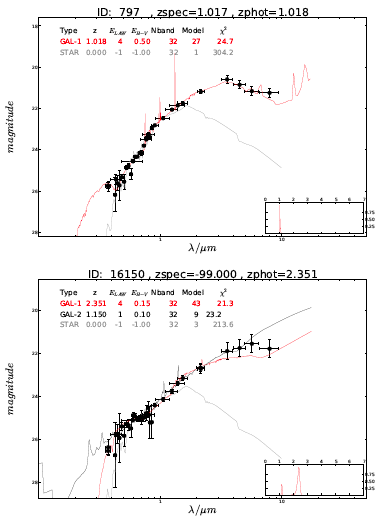}
 \caption{ Two examples of SED fitting for source 797 ( a normal
   galaxy) and source 16150 (an X-ray-detected AGN). The photometric
   points are shown in black. The red lines show the best-fitting
   template, and grey lines the best fitting star (the latter a poor
   fit for both objects shown). In the right panel, the black line
   shows the second-best template. Information about the
   templates---type, photo-z, extinction
   law, extinction value, number of bands,  model
   identification, and
   $\chi^2$ of the fit---is given in the legends.
   Inserts show
   $P(z)$ for the sources. \label{SED_fitting}}
\end{figure}

Tables~8 through 11 give homogeneously computed photo-z and related
data for
all sources detected in the area covered by CANDELS/GOODS-S, CDFS,
and ECDFS survey. For each source we make also available the redshift
probability distribution function $P(z)$\footnote{The redshift
  probability distribution function is derived directly from the
  $\chi^{2}$: $P(z)\propto \rm{exp}\left ( -\frac{\chi ^{2}(z)-\chi
      ^{2}_{\rm{min}}}{2} \right )$; $1\sigma$ is estimated from
  $\chi^2(z)-\chi^{2}_{\rm{min}}= 1\,(68\%)$, and $2.3\sigma$ is
  estimated from $\chi^2(z)-\chi^{2}_{\rm{min}}= 6.63\,(99\%)$}. With
these data it is possible to construct figures like the inserts in
Figure~\ref{SED_fitting}. Because of
the large size of the $P(z)$ files, we provide them
at the link \url{http://www.mpe.mpg.de/XraySurveys}. In
lieu of the full $P(z)$, the catalogs provide a
proxy in the form of the normalized integral of the main probability
distribution $P(z_{{p}})\equiv 100\times \int P(z)dz$ with the integral
  over the range $z_{{p}} \pm
  0.1(1+z_{{p}})$. A value close to 100 indicates that the
photo-z value is uniquely defined. Smaller values imply that a wide range
or multiple photo-z values are possible. Updated versions of the
catalogs and templates will be available under [Surveys] $>$ [CDFS] 
through the portal \url{http://www.mpe.mpg.de/XraySurveys}.

For the {\it Chandra} X-ray detections, the catalogs
also provide a new compilation of X-ray source
lists from the literature, the new optical/NIR/MIR associations, and
the corresponding photometry. Catalog descriptions and excerpts are
below. An entry of -99 indicates no data for that quantity. All coordinates are J2000.

\subsection{Cross ID reference table} 

 Table~\ref{crossid} gives cross-IDs and positions for all
  sources within
  each Area as identified  in Table~\ref{sets}. The table also
  indicates whether a source is a possible counterpart to an X-ray
  detection.

\begin{table*}\footnotesize
\begin{center}
\caption{Column description of the cross ID reference catalog}
\begin{tabular}{lll}
\toprule[1.5pt]
Column & Title & Description \\
\midrule
1        & [HSN2014] & Sequential number adopted in this work.\\
2-4     & $\rm{ID_{C}}$ , $\rm{R.A._{C}}$, $\rm{Dec._{C}}$ &  ID, right ascension and declination from the CANDELS-TFIT catalog (G13). \\
5-7     & $\rm{ID_{M}}$ , $\rm{R.A._{M}}$, $\rm{Dec._{M}}$& ID, right ascension and declination from the MUSYC catalog \citep{car10}.\\ 
8-10   & $\rm{ID_{T}}$, $\rm{R.A._{T}}$, $\rm{Dec._{T}}$ & ID, right ascension and declination from the TENIS catalog \citep{hsi12}.\\
11-13 & $\rm{ID_{S}}$, $\rm{R.A._{S}}$, $\rm{Dec._{S}}$ & ID, right ascension and declination from the SIMPLE catalog \citep{dam11}.\\
14-17  & $\rm{ID_{R13}}$, $\rm{R.A._{R13}}$, $\rm{Dec._{R13}}$, $\rm{PosErr_{R13}}$& ID, right ascension, declination and positional error from the R13 4Ms-CDFS catalog.\\
18-21&$\rm{ID_{X11}}$, $\rm{R.A._{X11}}$, $\rm{Dec._{X11}}$, $\rm{PosErr_{X11}}$ & ID, right ascension, declination and positional error from the X11 4Ms-CDFS catalog.\\
22-25 & $\rm{ID_{L05}}$, $\rm{R.A._{L05}}$, $\rm{Dec._{L05}}$, $\rm{PosErr_{L05}}$ & ID, right ascension, declination and positional error from the L05 250ks-ECDFS catalog. \\
26-29 & $\rm{ID_{V06}}$, $\rm{R.A._{V06}}$, $\rm{Dec._{V06}}$, $\rm{PosErr_{V06}}$ & ID, right ascension, declination and positional error from the V06 250ks-ECDFS catalog. \\
30      & Xflag & ``1" indicates that the source is the only possible counterpart to an X-ray source. \\
           &         & ``n" (2 or more) indicates that the source is one of the ``n"  possible counterparts \\
           &         & for a give X-ray source. ``-99" indicates that no X-ray counterpart are found. \\ 
31      & $p$   & Posterior value which indicates the reliability of the X-ray to optical/NIR/MIR association. \\ 
                     && (as defined in Section~\ref{sec:matching_method}) \\

\bottomrule[1.5pt]
\end{tabular}
\label{crossid}
\end{center}
\end{table*}

\subsection{X-ray source list in ECDFS}

\begin{table*}
\begin{center}
\caption{X-ray source list}
\begin{tabular}{ccccccccccccc}
\toprule[1.5pt]
[HSN2014]  &$\rm{ID_{R13}}$ &$\rm{ID_{X11}}$ &$\rm{ID_{L05}}$& $\rm{ID_{V06}}$ & $\rm{R.A._{x}}$ &$\rm{DEC._{x}}$ &$\rm{Flux_{s}}$ &$\rm{Flux_{h}}$ &$\rm{Flux_{f}}$  & log $L_{\rm{s}}$  &  log $L_{\rm{h}}$ &  log $L_{\rm{f}}$\\
(1)  &   (2)  & (3) &  (4)&    (5)&  (6) & (7)  & (8)  & (9) &(10)&(11)&(12)&(13)\\
\midrule
  125  &  343    &266    &-99    &-99    &53.079439    &-27.949429     &2.05E-16 &7.62E-16 &9.86E-16 &41.28 &41.85  &41.969 \\
  482    &6      &336    &-99    &-99      &53.103424    &-27.933357   &8.84E-16 &3.67E-15 &4.59E-15 &42.37 &42.99 &43.085\\
  47821 &-99    &-99    &527    &445   &53.251375    &-27.980556  &1.06E-15 &2.33E-15 &3.22E-15 &42.66 &43.00 &43.14\\ 
  50721 &-99    &-99    &32     &348    &52.842417    &-27.965417  &2.07E-15 &1.61E-14 &1.81E-14 &42.14 &43.03  &43.08\\

\bottomrule[1.5pt]
\end{tabular}
\label{xlist}
\end{center}
\end{table*}

 Table~\ref{xlist} gives the X-ray source list in Areas~1+2+3 with the position and flux information 
  from the available catalogs. Columns are as follows:\\
\noindent
 (1) [HSN2014]: Sequential number adopted in this work.\\
 (2) $\rm{ID_{R13}}$: ID from R13 catalog\\
 (3) $\rm{ID_{X11}}$: ID from X11 catalog\\
 (4) $\rm{ID_{L05}}$:  ID from L05 catalog \\
 (5) $\rm{ID_{V06}}$:  ID from V06 catalog\\
 (6) $\rm{R.A._{x}}$: Right Ascension of the X-ray source.\\
 (7) $\rm{DEC._{x}}$: Declination of the X-ray sources.\\
 (8) $\rm{Flux_{s}}$: Soft band X-ray flux ($\rm{erg~cm^{-2}~s^{-1} }$). \\
 (9) $\rm{Flux_{h}}$: Hard band X-ray flux. \\
 (10) $\rm{Flux_{f}}$: Full band X-rayflux. \\
 (11) log $L_{\rm{s}}$: Soft band X-ray luminosity ($\rm{erg~s^{-1} }$). \\
 (12) log $L_{\rm{h}}$: Hard band X-ray luminosity.\\
 (13) log $L_{\rm{f}}$: Full band X-ray luminosity.\\ 
 Note: From column (6) to (10), we chose the original X-ray data from, in order of priority, R13, X11, L05 and V06.

\subsection{Photometry of X-ray sources}

 Table~\ref{xphotometry} gives photometry 
   for all the possible counterparts to the X-ray sources.
  For the CANDELS area, this includes
  the TFIT photometry in the IBs as
  described in Section~\ref{optdata}. Columns are as follows:\\
\noindent
(1) [HSN2014]: Sequential number adopted in this work.\\
(2)-(5) XID:  ID from the four X-ray catalogs with the same order as Table~\ref{xlist} \\  
(6)Xflag: As described in Table~\ref{crossid} \\      
(7)$p$: As described in Table~\ref{crossid} \\  
(8)$\rm{R.A._{opt}}$: Right Ascension of the optical/NIR/MIR source. \\
(9)$\rm{Dec._{opt}}$: Declination of the optical/NIR/MIR source. \\
(10)-(109): AB magnitude and the associated uncertainty in each of possible bands (Table~\ref{photometry}).\\

\begin{table*}
\begin{center}
\caption{Photometry of X-ray sources}
\begin{tabular}{cccccccccccccc}
\toprule[1.5pt]
[HSN2014] &XID &xflag &p &$\rm{R.A._{opt}}$&$\rm{Dec._{opt}}$&$\rm{FUV_{m}}$  &$\rm{FUV_{e}}$&$\rm{NUV_{m}}$ & $\rm{NUV_{e}}$ &... &...&$\rm{IRAC4_{m}}$&$\rm{IRAC4_{e}}$ \\
(1)  &   (2)-(5)  & (6) &  (7)&    (8)&  (9) & (10)  &(11) &(12) &(13)&... &... &(108)&(109)\\
\midrule
  125  & ... &2   &0.98  &53.079489 &-27.948735 &-99.0 &-99.0 &-99.0 &-99.0 &... &... &19.888 &0.016\\
  482  & ... &1   &0.99    &53.103520 &-27.933323  &-99.0 &-99.0 &-99.0 &-99.0 &... &... &21.096 &0.03 \\
  47821& ... &2   &0.97  &53.252067  &-27.980645  &-99.0 &-99.0 &-99.0 &-99.0 &... &... &22.421 &0.18 \\
  50721& ... &2   & 1.0   &52.84249   &-27.965261  &-99.0 &-99.0 &-99.0 &-99.0 &... &... &19.24  &0.032\\

\bottomrule[1.5pt]
\end{tabular}
\label{xphotometry}
\end{center}
\end{table*}

\subsection{Redshift catalog}

\begin{table*}\footnotesize
\begin{center}
\caption{Redshift catalog}
\begin{tabular}{ccccccccccccccccc}
\toprule[1.5pt]
[HSN2014] &$\rm{R.A._{opt}}$&$\rm{Dec._{opt}}$& $z_{s}$ &$Q_{\rm{zs}}$  & $z_{p}$  &$1\sigma^{\rm{low}}$ &$1\sigma^{\rm{up}}$ &$3\sigma^{\rm{low}}$ &$3\sigma^{\rm{up}}$  &$P(z_{\rm{p}}$)&$z_{\rm{p}}2$ &$P(z_{\rm{p}}2$)&$N_{\rm{p}}$&Mod&xflag &$p$ \\
(1)  &   (2)  & (3) &  (4)&    (5)&  (6) & (7)  & (8)  & (9) &(10)&(11)&(12)&(13)&(14)&(15)&(16)&(17)\\
\midrule
13  & 53.093452 & -27.957135 & -99.0 & -99 & 3.2619 & 3.25 & 3.27 & 3.21 & 3.29 & 100.0 & -99.0 & 0.0 & 29 & 328&-99&-99 \\
14 & 53.104490 & -27.957068 & -99.0 & -99 & 2.1768 & 0.44 & 2.23 & 0.43 & 2.46 & 91.05 & 0.45 & 8.91 & 27 & 331&-99&-99 \\
15 & 53.088446 & -27.956996 & -99.0 & -99 & 3.0468 & 3.01 & 3.09 & 2.92 & 3.18 & 100.0 & -99.0 & 0.0 & 26 & 322&-99&-99 \\
16 & 53.104181 & -27.956592 & -99.0 & -99 & 3.1233 & 3.05 & 3.19 & 2.93 & 3.28 & 99.99 & -99.0 & 0.0 & 27 & 324&-99&-99 \\
125 & 53.079490 & -27.94874 & 0.619 & 0 & 0.6664 & 0.66 & 0.68 & 0.65 & 0.68 & 100.0 & -99.0 & 0.0 & 24 & 028&2&0.98 \\
135 & 53.142288 & -27.94447 & -99.0 & -99 & 2.6422 & 2.5 & 2.72 & 1.11 & 3.07 & 66.19 & 1.17 & 13.80 & 27 & 014&1&0.95 \\
\bottomrule[1.5pt]
\end{tabular}
\label{phz_cat}
\end{center}
\end{table*}

 Table~\ref{phz_cat} gives photo-z results for all sources detected in
  the CANDELS/GOODS-S, CDFS and ECDFS area. X-ray detections are
  flagged in the catalog. Columns are:\\
\noindent
 (1) [HSN2014]: Sequential number adopted in this work.\\
 (2) $\rm{R.A._{opt}}$: Right Ascension of the optical/NIR/MIR source.\\
 (3) $\rm{Dec._{opt}}$: Declination of the optical/NIR/MIR source.\\
 (4) $z_{\rm{s}}$: Spectroscopic redshift (N. Hathi, private communication). \\
 (5) $Q_{\rm{zs}}$: Quality of the spectroscopic redshift. (0=High, 1=Good, 2=Intermediate, 3=Poor).\\
 (6) $z_{\rm{p}}$: The photo-z value as defined by the minimum of $\chi^2$.\\
 (7) $1\sigma^{\rm{low}}$: Upper $1\sigma$ value of the photo-z.\\
 (8) $1\sigma^{\rm{up}}$: Lower $1\sigma$ value of the photo-z.\\
 (9) $3\sigma^{\rm{low}}$: Upper $2.3\sigma$ value of the photo-z.\\
 (10) $3\sigma^{\rm{up}}$: Lower $2.3\sigma$ value of the photo-z.\\
 (11) $P(z_{{p}})$: Normalized area under the curve $P(z)$, computed between $z_{\rm{p}} \pm 0.1(1+z_{{p}})$.\\
 (12) $z_{{p}}2$: The second solution in the photo-z, when the $P(z_{{p}2})$ is above 5. \\
 (13) $P(z_{{p}2})$: Normalized area under the curve $P(z)$, computed between $z_{{p2}} \pm 0.1(1+z_{{p2}})$.\\
 (14) $N_{\rm{p}}$:  Number of photometric points used in the fit.\\
 (15) Mod: Template number used for SED fitting. 1-48 are the
 templates from Lib-EXT; 101-130 are the templates from Lib-PT;
 201-230 are the  templates from S09; 301-331 are the templates from
 \citet{ilb09}, in the same order as the mentioned authors used.\\ 
(16) Xflag: As described in Table~\ref{crossid} \\      
(17) $p$: Posterior value which indicates the reliability of the X-ray
to Optical/NIR/MIR association (as defined in
Section~\ref{sec:matching_method}). \\

\section{Summary} \label{summary}

 The main product of this work is photometric redshifts for
  all sources detected in the CANDELS/GOODS-S, CDFS, and
  ECDFS area, a total of 105150 sources. This work has improved upon
  prior catalogs by G13, 
\citet{car10} and \citet{hsi12} by 
using the most up-to-date 
photometry and SED template libraries including separate libraries
for X-ray sources of different characteristics. Probabilities of
association between X-ray sources and optical/NIR/MIR sources are also
provided.

Our work has improved photo-z in the fields in the following ways: 

1. In the CANDELS area, we added the IB photometry from {\it Subaru}
\citep{car10} to the 
space-based photometric catalog of \citet{guo13}
using the same TFIT
parameters as in the official CANDELS catalog. The combined effect of
using IB photometry  to pinpoint 
emission lines in the objects and including lines in the templates 
gives excellent results, even for faint and high redshift sources
(Table~\ref{gal_phz_tab} and Table~\ref{xmix_phz_tab}). 

2. Using homogeneous data from the CANDELS/$H$-band, TENIS/$J\&K$ ,
MUSYC/$BVR$, and IRAC-3.6$\mu \mathrm{m}$-selected catalogs, we made
X-ray to multi-wavelength associations simultaneously by means of a
new, fast matching algorithm based on Bayesian statistics. This gave
$98\%$, $96\%$, and $94\%$ of X-ray sources with reliable
counterparts in Areas~1, ~2, and~3, respectively. Despite the new
technique and data,
all but 7 associations are consistent with those found earlier by
by X11. The 7 new associations come from
the deep, high-resolution CANDELS images and TENIS images that were not
available earlier. Different X-ray reduction procedures can change
the X-ray position by a few arcseconds. In crowded areas this may imply a
different X-ray to optical association. 

3. We demonstrated that the X-ray properties of sources need to be
taken into account when constructing the library of templates for
computing photo-z for such sources. More
specifically, the library defined by \citet{sal09,sal11} for the rare
X-ray bright sources detected in COSMOS is not representative of the
faint X-ray source population detected in the 
deeper 4Ms-CDFS. We therefore
defined a new of galaxy/AGN hybrids for the 4Ms survey
(Areas~1+2). In the 250ks 
survey (Area~3), where the X-ray data have a depth similar to
{\it Chandra}-COSMOS, the \citet{sal09} template library with the 
\citet{sal11} selection strategy works well.

\acknowledgments

We are grateful to the referee for constructive comments and to
Olivier Ilbert for the help with {\it LePhare}. This work
was supported by program number HST GO-12060 provided by NASA
through a grant from the Space Telescope Science Institute, which is
operated by the Association of Universities for Research in
Astronomy, Incorporated, under NASA contract NAS5-26555. We also
acknowledge the use of TOPCAT tool \citep{tay05}. This work is based in part on
observations made with the Spitzer Space Telescope, which is operated
by the Jet Propulsion Laboratory, California Institute of Technology
under a contract with NASA. Support for this work was provided by
NASA through an award issued by JPL/Caltech. M.\ Brusa acknowledges
support from the FP7 Career Integration Grant ``eEASy'' (CIG
321913). 

Facilities: {\it HST} (ACS/WFC3), {\it Spitzer} (IRAC), {\it
  Chandra}, {\it Subaru}, {\it GALEX}

\bibliography{lib}

\clearpage

\appendix

\section{AGN-Galaxy Hybrid Templates} \label{hybrids} 

We built hybrid templates by combining one of two AGN templates with
a set of normal galaxy templates in various proportions.
The Type~1 AGN template was derived from ``TQSO1'' of
\citet{pol07}. \citet{sal09} added a UV power-law to give
``pl-TQSO1.'' The Type~2 AGN template was ``QSO2'' unchanged from
\citet{pol07}.

\begin{figure}[h]
   \centering
    \includegraphics[width=0.40\textwidth]{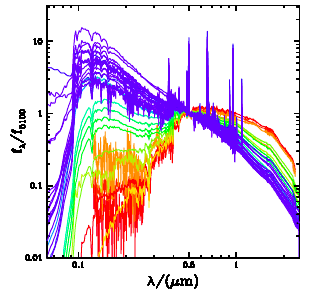}
    \caption{Galaxy templates from \citet{ben01}, color coded as a
      function of activity from the redder passive galaxies to the
      bluer strongly star forming objects. \label{bender_sed}} 
\end{figure}

\begin{figure}[h]
     \centering
      \includegraphics[width=\textwidth]{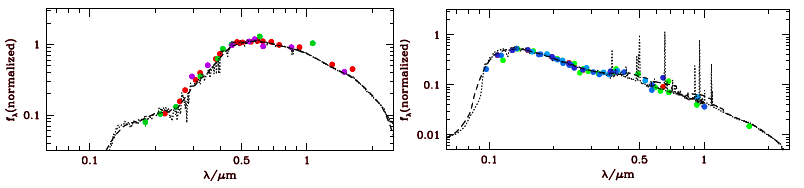}
 \caption{ Galaxy  templates in use for photometric
   redshift estimation. Colored symbols represent broad-band flux densities
   of individual galaxies with known spec-z. The short-dashed line
   shows the $R\sim1000$ spectral template best fitting the broadband
   flux densities after smoothing to the broad-band resolution. The smoothed
   template is shown as a long-dashed line.\label{SED-phz}}  
\end{figure}

The galaxy
templates were 32 semi-empirical ones from \citet{ben01} (see
Fig.~\ref{bender_sed}). These templates were constructed by
first sorting galaxies of known spec-z in the FORS Deep Field
\citep{app04,gab04} iteratively into 32 bins of similar spectral
shape. Broad-band fluxes from the $U$-band to the $K$-band of
typically ten galaxies at different redshifts were combined to obtain
one broadband template covering as wide a wavelength range as
possible. These broad-band empirical templates were fitted by a
combination of model spectral energy distributions from \citet{BC03}
and \citet{maraston05} and empirical spectra from \citet{noll04} to
obtain ``semi-empirical'' templates with spectral resolution
$R\sim1000$. The method covered  wavelengths
from $\sim$60~nm to 2.5~$\mu$m. Figure~\ref{SED-phz}
illustrates two of the templates in use.

\begin{figure}
     \centering
      \includegraphics[width=0.8\textwidth]{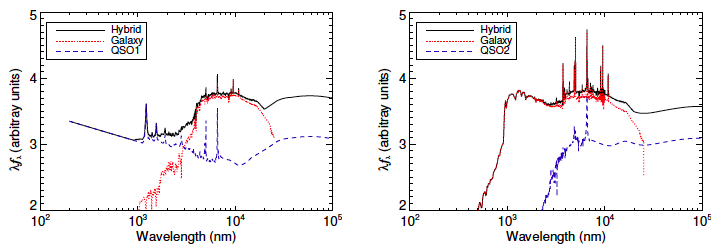}
   \caption{Two examples of hybrid SED templates. Red lines show the
     galaxy contribution, blue lines the AGN contribution, and black
     lines the sum. The left panel shows is a hybrid comprised of
     10\% Type~1 AGN and
     90\% galaxy. The right panel shows a hybrid with 30\%
     Type~2 QSO and 70\% starburst galaxy.\label{hybrid}}
\end{figure}

\begin{figure*}
      \centering
      \includegraphics[width=0.8\textwidth]{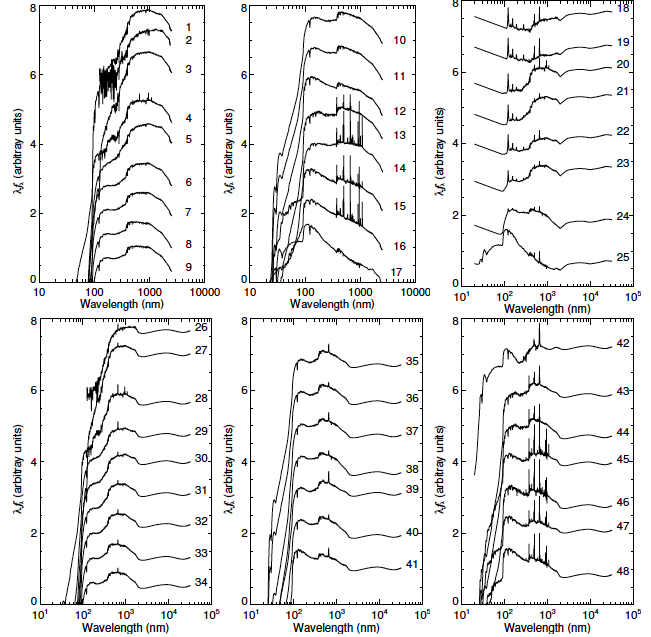}  
      \caption{SEDs for all templates in Lib-EXT.\label{SED-ext}} 
 \end{figure*}

\begin{figure*}
      \centering
      \includegraphics[width=0.8\textwidth]{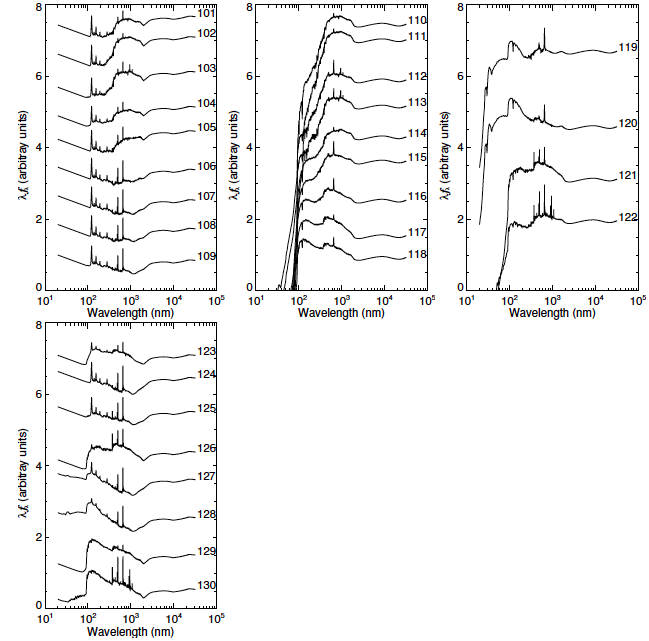}  
      \caption{SEDs for all templaes in Lib-PT.\label{SED-pt}} 
\end{figure*}

Making the hybrid templates followed the procedure of
\citet{sal09}. First we normalized both AGN and galaxy templates
at 5500~\AA, then combined them with the
AGN-to-galaxy ratio changing from 1:9 to 9:1. (See examples in
Fig.~\ref{hybrid}.)  In total, 576 hybrids were created this way.
We then randomly chose $25\%$ of spectroscopic X-ray sources
(52 extended sources and 62 point-like sources) to train the hybrids.
As Figure~\ref{hmag_spz} has shown, the training samples
are well distributed over the entire ranges of redshift and
magnitude. We treated the extended and point-like sources
separately, fixing the redshift at the spectroscopically defined
value and choosing the templates most frequently selected
to represent the training sources. After several iterations, we obtained the
libraries used for the extended sources (Lib-EXT: 31 hybrids + 17
galaxy templates, see Fig.~\ref{SED-ext}) and for the point-like
sources (Lib-PT: 30 hybrids, see Fig.~\ref{SED-pt}).

Table~\ref{lib_tab} lists the templates in Lib-EXT and
Lib-PT. Names with "-TQSO1-" or "-QSO2-" indicate the AGN component
used. The number following indicates
the fractional AGN contribution in the hybrid. For
example, the template ``s050-8-TQSO1-2'' contains $80\%$ galaxy
(s050) and $20\%$ of AGN (TQOS1) . The templates without TQSO1 and
QSO2 are pure galaxies with different levels of star formation. (See
\citealt{ben01} for details.)

\begin{table*}
\begin{center}
\caption{The list of AGN-galaxy hybrids in Lib-EXT and Lib-PT. \label{lib_tab}}
\begin{tabular}{cc|cc}
\toprule[1.5pt]
  \multicolumn{2}{c|}{Lib-EXT}&\multicolumn{2}{c}{Lib-PT} \\
  \midrule
  No. & Template & No. &Template \\
  \midrule
   1&mod-e &    101&e-8-TQSO1-2 \\
   2&manucci-sbc &   102&s010-9-TQSO1-1 \\
   3&mod-s010 &   103&s020-9-TQSO1-1 \\
   4&mod-s020 &   104&s050-8-TQSO1-2 \\
   5&mod-s030 &   105&sac-7-TQSO1-3 \\
   6&mod-s070 &   106&ec-3-TQSO1-7 \\
   7&mod-s090 &   107&sac-2-TQSO1-8 \\
   8&mod-s120 &   108&s010-3-TQSO1-7 \\
   9&mod-s150 &   109&s180-3-TQSO1-7 \\
   10&mod-s200 &   110&e-9-QSO2-1 \\
   11&mod-s400 &   111&s010-9-QSO2-1 \\
   12&mod-s500 &   112&s020-7-QSO2-3 \\
   13&mod-fdf4 &   113&s020-9-QSO2-1 \\
   14&mod-s210 &   114&s050-9-QSO2-1 \\
   15&mod-s670 &   115&s090-6-QSO2-4 \\
   16&mod-s700 &   116&s200-7-QSO2-3 \\
   17&mod-s800 &   117&s400-9-QSO2-1 \\
   18&ec-6-TQSO1-4 &   118&s500-8-QSO2-2 \\
   19&sac-5-TQSO1-5 &   119&s800-2-QSO2-8 \\
   20&s020-9-TQSO1-1 &   120&s800-5-QSO2-5 \\
   21&s030-9-TQSO1-1 &   121&fdf4-9-QSO2-1 \\
   22&s050-8-TQSO1-2 &   122&s230-5-QSO2-5 \\
   23&s070-9-TQSO1-1 &   123&s250-8-TQSO1-2 \\
   24&s250-9-TQSO1-1 &   124&s250-1-TQSO1-9 \\
   25&s800-8-TQSO1-2 &   125&fdf4-4-TQSO1-6 \\
   26&sac-9-QSO2-1 &    126&fdf4-9-TQSO1-1 \\
   27&s010-9-QSO2-1 &   127&s800-2-TQSO1-8 \\
   28&s020-9-QSO2-1 &   128&s800-4-TQSO1-6 \\
   29&s050-8-QSO2-2 &   129&s500-9-TQSO1-1 \\
   30&s050-9-QSO2-1 &   130&s670-9-TQSO1-1 \\
   31&s070-9-QSO2-1 & & \\
   32&s090-9-QSO2-1 & & \\
   33&s120-9-QSO2-1 & & \\
   34&s180-9-QSO2-1 & & \\
   35&s200-8-QSO2-2 & & \\
   36&s200-9-QSO2-1 & & \\
   37&s250-8-QSO2-2 & & \\
   38&s250-9-QSO2-1 & & \\
   39&s400-7-QSO2-3 & & \\
   40&s400-9-QSO2-1 & & \\
   41&s500-8-QSO2-2 & & \\
   42&s800-1-QSO2-9 & & \\
   43&fdf4-7-QSO2-3 & & \\
   44&fdf4-9-QSO2-1 & & \\
   45&s230-8-QSO2-2 & & \\
   46&s650-9-QSO2-1 & & \\
   47&s670-6-QSO2-4 & & \\
   48&s670-9-QSO2-1 & & \\
\bottomrule[1.5pt]
\end{tabular}
\end{center}
\end{table*}

\end{document}